# A Multinomial Probit Model with Choquet Integral and Attribute Cut-offs

Subodh Dubey[1], Oded Cats[1], Serge Hoogendoorn[1], Prateek Bansal[2*]

[1]Department of Transport & Planning, Delft University of Technology, The Netherlands

[2]Department of Civil and Environmental Engineering, Imperial College London, The United Kingdom

*Corresponding author (pb422@cornell.edu)

**Abstract**

Several non-linear functions and machine learning methods have been developed for flexible specification of the systematic utility in discrete choice models. However, they lack interpretability, do not ensure monotonicity conditions, and restrict substitution patterns. We address the first two challenges by modelling the systematic utility using the Choquet Integral (CI) function and the last one by embedding CI into the multinomial probit (MNP) choice probability kernel. We also extend the MNP-CI model to account for attribute cut-offs that enable a modeller to approximately mimic the semi-compensatory behaviour using the traditional choice experiment data. The MNP-CI model is estimated using a constrained maximum likelihood approach, and its statistical properties are validated through a comprehensive Monte Carlo study. The CI-based choice model is empirically advantageous as it captures interaction effects while maintaining monotonicity. It also provides information on the complementarity between pairs of attributes coupled with their importance ranking as a by-product of the estimation. These insights could potentially assist policymakers in making policies to improve the preference level for an alternative. These advantages of the MNP-CI model with attribute cut-offs are illustrated in an empirical application to understand New Yorkers' preferences towards mobility-on-demand services.

# 1. Introduction

Eliciting individual-level decisions is of interest in multiple disciplines, such as transportation, economics, environment, ecology, and health, among others. Discrete choice models relying on random utility maximisation (RUM) theory are still workhorse models in these disciplines (McFadden, 1973; Train, 2009). RUM-based models represent the preferences of decision-makers through latent stochastic utility. Most applications assume that the indirect utility consists of a linear-in-parameters systematic utility and additive stochastic unobserved term, with a few instances of the multiplicative stochastic term (Fosgerau and Bierlaire, 2009). Several recent studies have adopted machine learning techniques for flexible representation and data-driven learning of the systematic utility (see Aboutaleb et al. 2021; Hillel et al. 2021; Van Cranenburgh et al., 2021 for literature review). These techniques include kernel smoothing (Bansal et al., 2019), deep learning architectures (Ortelli et al., 2021; Sifringer et al., 2020; Wang et al., 2021; and Wong and Farooq, 2021), and automatic relevance determination (Rodrigues et al., 2020). We identify three main shortcomings of the existing studies. First, the flexible specifications do not ensure the monotonicity of the utility function relative to attributes like cost, which is a necessary condition for a demand function to be valid[1]. Second, most theory-driven machine learning studies claim that they do not compromise interpretability, but their notion of interpretability is no longer associated with the behavioural or physical interpretation of model parameters. For instance, Wang et al. (2021) define interpretation quality metrics based on the learned choice probability function, which can only be measured in simulation studies. Third, previous studies with flexible systematic utility capture restrictive substitution effects among alternatives because they rely on multinomial or nested logit choice probability kernel.

This study addresses the first two limitations by specifying the systematic utility using the Choquet Integral (CI) function and the last one by embedding it into the multinomial probit (MNP) choice probability kernel (i.e., MNP-CI model henceforth). The CI function nests different aggregation functions – weighted means (i.e., linear-in-parameters), ordered weighted averaging functions, minimum, maximum, and order statistics (Grabisch and Labreuche, 2010), offers a systematic way to capture all possible interactions between attributes, and ensures monotonicity in terms of the number of attributes[2] and attribute values. Additionally, the MNP kernel can represent flexible substitution patterns with relatively less computational complexity than a logit kernel. The MNP kernel is also computationally feasible for joint modelling of several choice dimensions and social effects due to the elegant properties of the Gaussian distribution (Astroza et al., 2018; Vinayak et al., 2018). Therefore, demonstrating the application of the CI function embedded into the MNP kernel is practically more relevant.

[1] The specifications similar to Sifringer et al. (2020) could ensure monotonicity by decomposing the utility into flexible and linear-in-parameter components, and including the attributes with directional effect in the latter component. However, since most attributes have a directional effect on choices in stated preference studies, the resulting utility would be driven by the linear-in-parameters part.

CI has been a popular aggregation operator in multi-attribute decision making and preference learning literature (Alfonso, 2013; Grabisch, 1996). Most studies are concerned with improving the prediction performance (Tehrani et al., 2012; Sobrie et al., 2015; Cano et al., 2019). We identify two main limitations of the literature on CI applications and address them in this study. First, many studies have explored the application of CI in RUM-based logit models (Aggarwal, 2018; Aggarwal, 2019; Aggarwal, 2020; Büyüközkan et al., 2018; Demirel et al., 2017), but fail to develop econometrically-sound estimators and restrict substitution effects. In contrast, accommodating unrestrictive substitution effects is straightforward in the MNP-CI model. We also estimate the MNP-CI model using a constrained maximum likelihood estimator, and establish its statistical properties (e.g., bias and coverage probability) in a Monte Carlo study. Second, the normalisation of attributes across alternatives in traditional CI-based choice models is not feasible in the case of unbalanced attribute configuration (i.e., when different subsets of the attributes are applicable for different alternatives). We resolve this limitation by extending the MNP-CI model to an alternative-specific MNP-CI specification (i.e., analogous to the MNP specification with alternative-specific marginal utilities).

We advance the alternative-specific MNP-CI specification to account for the attribute-level cut-offs or constraints. The MNP-CI model with attribute cut-offs is closely related to one-stage semi-compensatory models (Ding et al., 2012; Elrod et al., 2004; Martínez et al., 2009; Swait 2001; Truong et al., 2015). Swait (2001) formulated the first one-stage semi-compensatory model as the reduced form approximation of the Manski (1977)'s two-stage model – choice set formation based on non-compensatory screening process (e.g., elimination-by-aspects and conjunctive rules) in the first stage, followed by evaluation of the remaining alternatives based on the compensatory decision rule in the second stage (Gilbride and Allenby, 2004; Cantillo and de Dios Ortúzar, 2005; Kaplan et al., 2012). Instead of posing *hard constraints* on the elimination of alternatives in the first stage of Manski's approach, Swait's one-stage model puts *soft constraints* by allowing the decision-maker to violate cut-off rules at the cost of utility penalisation, hence allowing to choose the alternative with attribute cut-off violation if it still has the highest utility even after penalisation (Rashedi and Nurul Habib, 2020). Such one-stage models are empirically attractive because they could approximately mimic the semi-compensatory behaviour by adjusting the systematic utility specification within the traditional RUM framework[3].

[2] Monotonicity in the number of attributes implies that the attribute addition should always increase the informational power (e.g., R-square).

We highlight some key differences between the MNP-CI model and the existing one-stage semi-compensatory models. The existing models rely on adding a penalty function in the utility to regulate the overall utility level based on attribute cut-off violations. However, the use of fuzzy-membership functions to model attribute cut-offs in MNP-CI obviates the need for penalty functions. Directly specifying cut-offs on attributes could be possible due to the normalisation of attributes and monotonicity constraints in the estimation of MNP-CI. In contrast, such direct attribute-cut-off specifications in the traditional MNP with linear-in-parameters utility often leads to numerical issues due to the unconstrained nature of the likelihood maximisation problem.

The existing semi-compensatory models are also subject to few shortcomings. Swait (2001) directly asked attribute cut-offs from the respondents, which could be susceptible to self-reporting bias. Martínez et al. (2009) addressed this limitation by endogenously estimating attribute cut-offs (known as constrained multinomial logit (CMNL) model)[4], but their estimator relies on solving a rather complex fixed-point problem with little evidence regarding its finite sample properties (see Section 2.5 for details). Moreover, existing one-stage semi-compensatory models consider logit kernel. We address these limitations in this study. Whereas the MNP kernel in the proposed model leads to unrestricted substitution effects, the constrained maximum likelihood estimator of MNP-CI can endogenously estimate attribute cut-offs and has valid statistical properties.

The contribution of this study is thus three-fold. First, a nonlinear additive functional form of systematic utility is specified using the CI function to capture interaction effects between attributes with strict monotonicity. Second, the CI-based choice model is extended to capture alternative-specific attribute importance and complementarity. The semi-compensatory behaviour is accounted by the means of endogenous attribute cut-offs. Third, the CI-based systematic utility is embedded in the MNP choice probability kernel to capture unconstrained substitution patterns. A constrained maximum likelihood estimator is developed for the proposed model, which incorporates constraints to maintain monotonicity requirements arising from the CI function. In addition to a Monte Carlo study, the practical relevance of the model is demonstrated in an empirical study to understand the preferences of New Yorkers for mobility-on-demand services. This paper thus makes advancements in three strands of the literature: 1) flexible specification of the systematic utility; 2) multi-attribute decision making and preference learning using the CI function; and 3) One-stage semi-compensatory behaviour modelling.

[3] Elrod et al. (2004) illustrated how various non-compensatory rules can be modelled within the RUM framework by specifying the systematic utility using a general nonrectangular hyperbola.
[4] Bierlaire et al. (2010) demonstrated that the CMNL model should be considered as a semi-compensatory model on its own because it poorly approximates Manski's two-stage framework.

The remainder of this paper is structured as follows. Section 2 provides a detailed discussion on the properties of the CI function, our modelling extensions, the estimation procedure, and the advantages of the proposed model. Section 3 details the simulation set up, evaluates statistical properties of the estimator and demonstrates the superiority of the proposed model over the traditional MNP model with linear-in-parameter systematic utility. Section 4 uses an empirical example to illustrate how the proposed model can offer interesting insights into the behaviour of a decision-maker. Conclusions and future work are summarised in Section 5.

## 2. Choquet Integral based Random Utility Choice Model

### 2.1 Properties of Choquet Integral

CI is a fuzzy integral based on fuzzy measures, which provides an elegant way to capture all possible interactions between attributes. For instance, CI allows the analyst to explicitly capture complementarity between attributes which may help explain the outcome (choices) more accurately. In mathematical terms, if the fuzzy measure $\mu(k)$ represents the informational power of attribute $k$, then the complementarity of two attributes implies that $\left(\mu(1,2) > \mu(1) + \mu(2)\right)$. The CI function also ensures monotonicity while allowing for flexible interactions between attributes. This characteristic of the CI function is critical because arbitrary flexible interactions between attributes generally lead to a non-monotonic utility function (Elrod et al., 2004).

CI ensures monotonicity while capturing attribute interactions through fuzzy measures. A discrete fuzzy measure allows one to assign importance to all possible combinations of attributes. In mathematical terms, one can define discrete fuzzy measures as follows:

$$\mu(\phi)=0 \quad \mu(\boldsymbol{X})=1 \quad \mu(\boldsymbol{A})\le\mu(\boldsymbol{B});\boldsymbol{A}\subseteq\boldsymbol{B}\subseteq\boldsymbol{X} \quad 0\le\mu(.)\le 1$$

where $\boldsymbol{A}$ and $\boldsymbol{B}$ are sets of attributes, $\phi$ represents the null set, and $\boldsymbol{X}$ is the set of all attributes. The fuzzy measures are monotonic in the number of attributes by definition because adding an attribute to an existing set does not decrease the importance of the new coalition. Normalisation of attributes before passing through fuzzy measures ensures monotonicity in attribute values (see Section 2.2 for details). We can write the CI with respect to a discrete fuzzy measure as follows:

$$CI = \sum_{g=1}^{G} h\left(x_{\pi_g}\right)\left(\mu\left(A_g\right)-\mu\left(A_{g-1}\right)\right)$$

where $A_g$ is the set of cardinality $g$ formed using permutation of attributes $(x)$, $g\in\{1,2,...,G\}$ is the index for attributes, and

$$h\left(x_{\pi_g}\right)\rightarrow h\left(x_{\pi_1}\right)\ge h\left(x_{\pi_2}\right)\ge ...\ge h\left(x_{\pi_G}\right)\ge 0$$
$$A_G=\{x_1,x_2,...,x_G\}$$

The function $h(.)$ represents the numerical value of attributes $(x)$ in a descending order. An example of the CI computation is provided in the online appendix S.1. There are two important points to observe. First, the number of fuzzy measures is a function of the number of attributes, i.e. the number of fuzzy measures is $2^G$, two of which are the null set and the complete set. Second, the term $\left(\mu\left(A_g\right)-\mu\left(A_{g-1}\right)\right)$ can be interpreted as the additional information that attribute $x_g$ offers in decision-making. This information can be used to interpret CI as a representation of an information processing strategy adopted by the decision-maker. One way to interpret CI-based decision-making process could be that individuals first pick the attribute that provides the maximum amount of information ( $h\left(x_{\pi_g}\right)\rightarrow h\left(x_{\pi_1}\right)\ge h\left(x_{\pi_2}\right)\ge ...\ge h\left(x_{\pi_G}\right)\ge 0$ ) while making the choice and assess its value by multiplying it with the corresponding fuzzy measure. Subsequently, the next attribute (in a decreasing order of amount of information offered) is selected and its additional contribution is assessed with ($\left(\mu\left(A_g\right)-\mu\left(A_{g-1}\right)\right)*x_g$). This procedure is followed until all attributes are parsed through. Of course, this is a mathematical interpretation of CI and may not correspond exactly to the underlying decision-behaviour mechanism.

With the use of examples in the online appendix S.2, we illustrate how CI can approximate various aggregation functions ranging from weighted sum, ordered sum, minimum or maximum of attributes. With these examples, we aim to convincingly argue for the candidacy of CI as a flexible and monotonic aggregation function in RUM framework. For a detailed discussion on CI, readers are referred to Tehrani et. al. (2012). In the next subsection, we illustrate how linear additive utility specification can be replaced with CI in the MNP model.

### 2.2 Multinomial Probit Choice Model with Choquet Integral (MNP-CI)

Traditional RUM-based discrete choice models use a weighted sum (WS) aggregation function to represent the systematic part of the indirect utility. In this section, we replace the WS with CI while retaining the stochastic part of the indirect utility function as the normally distributed random variable. For brevity, we refer to the MNP model with the CI function as MNP-CI and to the MNP model with WS function as MNP-WS.

In MNP-WS, the indirect utility of an individual $(n)$ from choosing alternative $i \in \{1, 2, ..., I\}$ as a function of attributes $g \in \{1, 2, ..., G\}$ is defined in Eq. 1 (suppresing individual-level subscript for notational simplicity):

$$U_i = v_i + \varepsilon_i = \boldsymbol{\beta}' \boldsymbol{x}_i + \varepsilon_i \tag{1}$$

where $\boldsymbol{x}_i$ is a $(G \times 1)$ vector of exogenous variables, $\boldsymbol{\beta}$ is the corresponding $(G \times 1)$ vector of marginal utilities, and $\varepsilon_i$ is a normally-distributed idiosyncratic error term. We replace the observed part of the utility ($v_i$) in the MNP-WS with CI and rewrite Eq. 1 as follows:

$$U_i = CI_i + \varepsilon_i$$

where

$$
\begin{aligned}
&CI_i = \sum_{g=1}^{G} h\left(x^i_{\pi_{N_g}}\right)\left(\mu\left(A^i_g\right) - \mu\left(A^i_{g-1}\right)\right) \\
&\text{where } A_k \text{ is the set of cardinality } k \text{ formed using permutation of } x \\
&h\left(x^i_{\pi_{N_g}}\right) \rightarrow 0 \le \left[h\left(x^i_{\pi_{N_1}}\right) \ge h\left(x^i_{\pi_{N_2}}\right) \ge \dots \ge h\left(x^i_{\pi_{N_G}}\right)\right] \le 1 \\
&A^i_G = \left\{x^i_{N_1}, x^i_{N_2}, \dots, x^i_{N_G}\right\}
\end{aligned}
\tag{2}
$$

In Eq.2, the function $h(.)$ is applied on the normalised attribute values. Note that $x^i_{N_g}$ and $x^i_g$ are normalised and un-normalised attribute values of attribute $g$ for alternative $i$. Further, readers will note that the calculation of CI involves the same set of fuzzy measures ($\mu$) for all the alternatives, and therefore, does not have any alternative-specific subscript. This specification is similar to the choice models with generic marginal utilities across alternatives. We extend the MNP-CI to accommodate alternative-specific fuzzy measures (by replacing $\mu$ with $\mu_i$) in Section 2.4.1. At this point, there are two additional conditions that need to be ensured in MNP-CI.

*First*, the attribute values are normalised between 0 and 1 in CI computation with 0 and 1 indicating the lowest and the highest amount of information provided by an attribute, respectively. Such rescaling ensures monotonicity in terms of attribute values. Rescaling the attribute values between 0 and 1 also offers additional stability during numerical optimisation because both parameters (fuzzy measures) and explanatory variables are on the same numerical scale.

The normalisation is performed by using the range of attributes across all available alternatives as illustrated below. Let $\psi\left(x_g\right) = \left\{x^1_g, x^2_g, \dots, x^I_g\right\}$ be the collection of $g^{th}$ attribute values across all alternatives. For attributes with positive effect on utility and choice probability (higher the value, better the attribute), the corresponding normalised value can be obtained as follows:

$$
x^i_{N_g} = \frac{x^i_g - \min\left(\psi\left(x_g\right)\right)}{\max\left(\psi\left(x_g\right)\right) - \min\left(\psi\left(x_g\right)\right)} \tag{3}
$$

Similarly, for attributes with a negative effect on utility (the lower the value, the better the attribute), the corresponding normalised value can be obtained as follows:

$$x_{N_g}^i = \frac{\max\left(\psi\left(x_g\right)\right) - x_g^i}{\max\left(\psi\left(x_g\right)\right) - \min\left(\psi\left(x_g\right)\right)} \tag{4}$$

Such normalisations ensure that the rescaled values are always a function of available alternatives. It can to some extent help avoid the independence of irrelevant alternatives (IIA) issue in the absence of a non-IID error structure.

*Second*, to ensure that $\mu(\boldsymbol{X}) = 1 \quad \mu(\boldsymbol{A}) \leq \mu(\boldsymbol{B}); \boldsymbol{A} \subseteq \boldsymbol{B} \subseteq \boldsymbol{X} \quad 0 \leq \mu(.) \leq 1$, we write constraints using Möbius transformation as the transformed space has one-to-one mapping with fuzzy measures:

$$\sum_{H \subseteq A_G} m(H) = 1; \quad \text{where } A_G = \{x_1, x_2, ..., x_G\}$$

$$\sum_{H \subseteq A_G \setminus g} m(H \cup k) \geq 0 \;\; \forall g \subseteq A_G, \forall k \subseteq A_G; \tag{5}$$

where $A_G \setminus g$ represents the collection of all attributes except the $g^{th}$ attribute

$\cup$ represents the union of two sets

$m(.)$ is the Möbius representation of $\mu(.)$ and one-to-one mapping between them is as follows:

$$m(H) = \sum_{F \subseteq H} (-1)^{|H \setminus F|} \mu(F)$$

$$\mu(F) = \sum_{H \subseteq F} m(H)$$

Thus, after estimation of Möbius parameters, one can derive the fuzzy-measure $\mu(.)$ from the estimated Möbius parameters $m(.)$ using the above equation. While the fuzzy measures are constrained between 0 and 1, Möbius parameters, except singleton elements, are unconstrained. An example in Section S.3 of the online appendix illustrates a mapping between Möbius parameters and fuzzy measures.

We also emphasize that the number of CI-specific parameters in the generic CI function depends on the number of attributes (rather than the number of alternatives). In other words, the computational challenges associated with the large choice sets in the MNP-CI model would be the same as those in the MNP model with linear-in-parameter utility. Specifically, the dimensionality of the integrals (i.e., multivariate normal cumulative density function) in the MNP choice probability kernel is one less than the number of alternatives. Nevertheless, advancements in quasi-Monte-Carlo, quadrature, and other analytical approximation methods have enabled efficient computations of high-dimensional integrals in the case of large choice sets (Bansal et al., 2021; Bhat, 2018).

In sum, unlike the MNP-WS model, the estimation of MNP-CI requires solving a constrained optimisation problem with a set of equality and inequality constraints. The addition of constraints means that the typically used Broyden–Fletcher–Goldfarb–Shanno (BFGS) algorithm (Fletcher, 2013) can no longer be used for the loglikelihood maximisation. Therefore, we use the sequential least-square programming (SLSQP) algorithm to solve the constrained loglikelihood maximisation problem of MNP-CI. Readers are referred to Nocedal et al. (2006, page 529-562) for a detailed discussion on the SLSQP algorithm. We use the SLSQP algorithm's off-the-shelf implementation in Python's Scipy package. A detailed description of the MNP-CI formulation and estimation is provided in Appendix A.1. This section illustrates the changes due to the replacement of the WS with CI, the normalisation of attributes, and the constrained likelihood maximisation problem.

### 2.3 Inferences from a Choice Model with Choquet Integral

Whereas Möbius parameters are direct output of the estimation, several important metrics can be derived after transforming them into fuzzy measures. We discuss two such metrics – Shapley value and interaction indices, which provide further insights into decision-making process. Readers are referred to Beliakov et al. (2016, chapter 4) for a detailed discussion on such metrics.

The Shapley value of an attribute is expressed as follows:

$$S(g) = \sum_{A \subset G \setminus g} \frac{Fact\left(|\boldsymbol{X}| - |A| - 1\right) Fact\left(|A|\right)}{Fact\left(|\boldsymbol{X}|\right)} \left[\mu\left(A \cup \{g\}\right) - \mu(A)\right] \quad 0 \le S(g) \le 1, \tag{6}$$

where $Fact(.)$ represents the factorial, $|.|$ indicates the cardinality of the set and $\boldsymbol{X} = \{x_1, x_2, ..., x_G\}$ is the set of all attributes. The Shapley value is interpreted as the average marginal contribution of an attribute $g$ in all coalitions. Intuitively, Eq. 6 provides the sum of scaled (multiplication by a factor) difference between fuzzy measures of sets with and without attribute $g$. In other words, the Shapley value basically aggregates all additional worth of attribute $g$ as represented by $\left[\mu\left(A \cup \{g\}\right) - \mu(A)\right]$.

While the Shapley value is informative, it is not sufficient to describe the entire effect of an attribute on the choice outcome because it does not capture the importance of the attribute's interaction with other attributes in explaining the choice outcome. However, interaction indices can address this limitation of the Shapley value.

The interaction index – a pair-wise value (which represents if two attributes are complementary or not) can be obtained as follows:

$$I(qw) = \sum_{A \subset G \setminus \{q,w\}} \frac{Fact(|\boldsymbol{X}| - |A| - 2)\, Fact(|A|)}{Fact(|\boldsymbol{X}| - 1)} \left[ \mu(A \cup \{q,w\}) - \mu(A \cup \{q\}) - \mu(A \cup \{w\}) + \mu(A) \right], \quad (7)$$

$$-1 \le I(qw) \le 1$$

Similar to Eq. 6, Eq. 7 essentially provides the sum of scaled differences between fuzzy measures of sets with and without the pair of attributes. A positive value of interaction index indicates a complementary relation (positive interaction) between two attributes and a negative value suggests otherwise.

The interaction index can also be obtained for a group of more than two attributes (set *B*) with the help of Eq. 8:

$$I(B) = \sum_{A \subset G \setminus \{B\}} \frac{Fact(|\boldsymbol{X}| - |A| - |B|)\, Fact(|A|)}{Fact(|\boldsymbol{X}| - |B| + 1)} \sum_{C \subseteq B} (-1)^{|B \setminus C|} \mu(A \cup C) \quad (8)$$

Grabisch and Roubens (2000) provide a policy-relevant interpretation of both Shapley value and interaction index in quantifying the effect of an attribute on the overall choice process – "*A positive value of interaction index implies a conjunctive behaviour between the pair of attributes. This means that the simultaneous satisfaction of both the attributes is significant for the final choice. On the other hand, a negative value implies a disjunctive behaviour, which means that the satisfaction based on either of attributes has a substantial impact on the final choice. Finally, the Shapley value acts as a weight vector in a weighted arithmetic mean, i.e. it represents the linear part of Choquet integral.*"

We illustrate the interpretation of both metrics in a mode choice context. Let us consider a travel mode choice scenario with three attributes namely price, comfort, and out-of-vehicle travel time (OVTT). Further, we assume that the Shapley value and interaction indices for the attribute pairs are as follow:

$$S(\text{Price}) = 0.45 \qquad , S(\text{Comfort}) = 0.25 \qquad , \text{and } S(\text{OVTT}) = 0.30$$
$$I(\text{Price,Comfort}) = -0.25, I(\text{Price,OVTT}) = 0.15, \text{ and } I(\text{Comfort,OVTT}) = 0.12$$

If we only consider Shapley values in isolation, then one may make the conclusion that price is the most important attribute for the traveller when making a travel mode choice, followed by OVTT and comfort. In other words, if one wishes to improve the share of a travel mode, then lowering the price followed by improving OVTT and comfort is likely to yield the best results. However, when we analyse interaction indices along with Shapley values, we observe that price and OVTT exhibit a complementary behaviour, i.e. the travel mode needs to score lower on both of these attributes (as they cause disutility) in order to be chosen. However, if a decision-maker chooses a travel mode based on price and comfort, the travel mode needs to score lower on price or substantially higher on comfort (due to a low Shapley value) in order to be chosen. This example demonstrates that Shapley values alone (individual ranking of attributes) are informative but not adequate, and interaction indices play an imperative role in making policy-relevant recommendations.

The above discussion suggests that there could be a structured way to identify important attributes (to increase the market share of an alternative) based on Shapley values and interaction indices. The analyst can focus on the attribute with the highest Shapley value and can find the corresponding complementary pair with the highest interaction index value. Any improvement in both attributes simultaneously is likely to improve the share of an alternative substantially. This process can be viewed as equivalent to a strategy where one may evaluate the elasticity value of both attributes individually and simultaneously to identify which yields maximum improvement in market share of an alternative.

## 2.4 Extensions of the Choice Model with Choquet Integral

In addition to operationalising CI in the MNP framework, we propose two extensions of MNP-CI. Readers are referred to Appendix A.1.1 for a generalised formulation of MNP-CI with these extensions.

### *2.4.1 Alternative-specific Choquet Integral*

Analogous to the MNP specification with alternative-specific marginal utilities, we extend the MNP-CI to an alternative-specific MNP-CI specification where different subsets of the attributes could be used for different alternatives. It is worth noting that the alternative-specific MNP-CI obviates several behavioural constrains such as the same ranking of attribute importance for all alternatives. For instance, there is no reason to assume that individuals attach the same importance to the price across all alternatives in the presence of brand loyalty. Thus, relaxing this assumption in alternative-specific MNP-CI allows the analyst to uncover important alternative-specific attribute ranking and complementary pairs of attributes. Unbalance datasets (i.e., when different subsets of the attributes are applicable for different alternatives) can also be easily handled using this specification.

*2.4.2 Choquet Integral with Attribute Cut-offs*

We explicitly incorporate and endogenously estimate attribute cut-offs to account for semi-compensatory behaviour of decision-makers. We parametrise attribute cut-offs with socio-demographic characteristics of decision-makers to inherently capture heterogeneity in preferences due to adoption of different attribute cut-off.

Since CI requires the analyst to rescale attribute values between 0 and 1, we can directly use fuzzy membership functions (e.g., triangular, sinusoidal, and trapezoidal) to specify attribute cut-offs. The selection of a membership function depends on the perception of the attribute. For instance, cut-off for attributes with negative marginal utility such as travel time and cost in travel model choice can be represented by the following half-triangular membership function:

$$x_N = \begin{cases} 1 & x \le a \\ \dfrac{b-x}{b-a} & a < x \le b \\ 0 & x > b \end{cases} \tag{9}$$

To illustrate the above equation, Figure 1 shows an example of how the membership function value changes with the change in actual travel time in a mode choice scenario. Figure 1 shows that the disutility of travel time remains constant below 10 minutes. The upper limit of 25 minutes indicates that all travel time values above 25 minutes offer similar level of disutility to the traveller as of 25 minutes. The linear change is similar to a regular MNP-CI normalisation with minimum 10 minutes and maximum 25 minutes, as illustrated in Eq. 4.

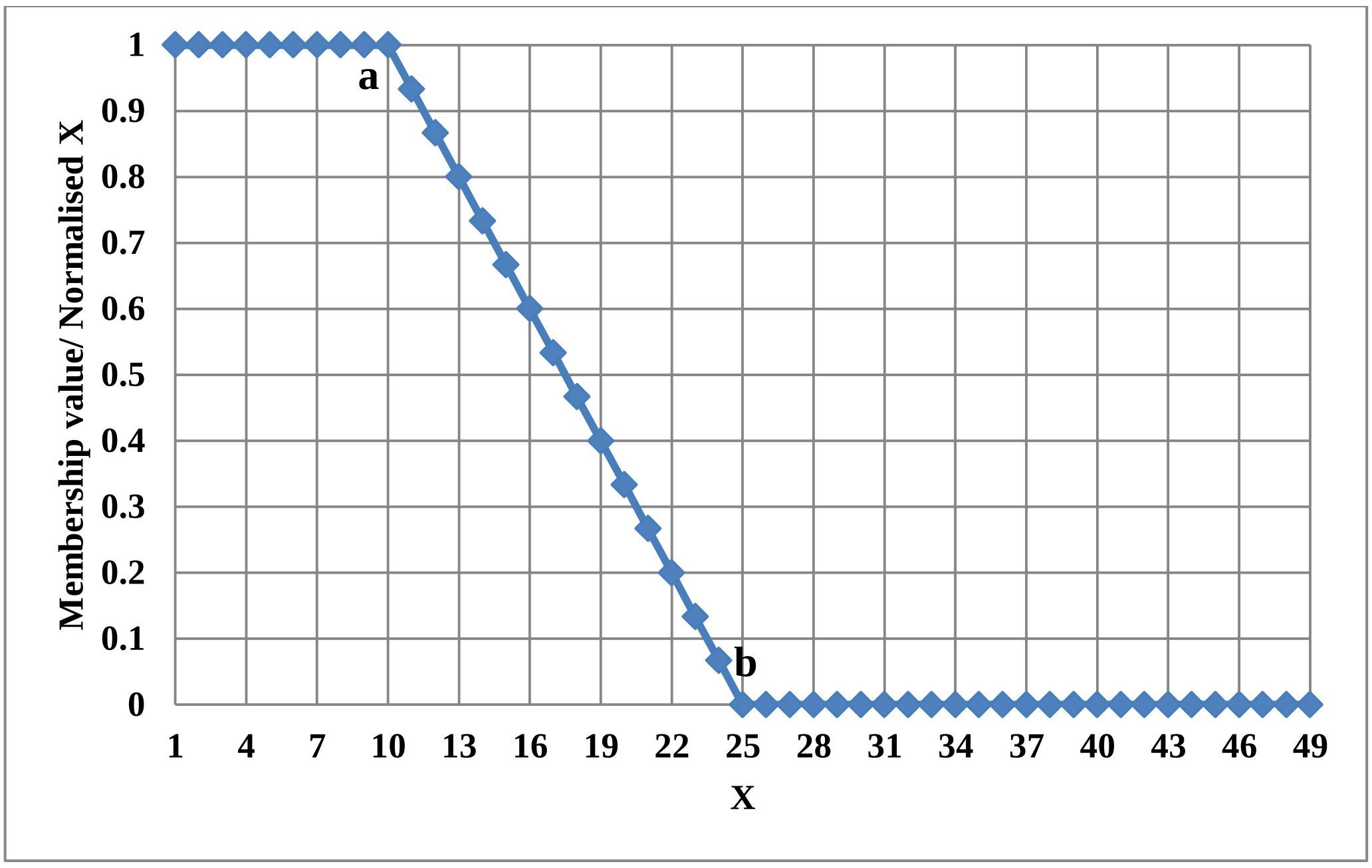

**Figure 1:** Two-point cut-off graph for attributes with negative marginal utility (Half-triangular function)

The application of these membership functions allows us to endogenously determine preference ranges for an attribute independently for each alternative and variation across different groups can be captured by parameterising kink points as a function of demographic characteristics. The kink point parameters are estimated during the estimation (loglikelihood maximisation) process.

In this attribute cut-off framework, behaviour remains compensatory in a certain range of attributes, but it becomes semi-compensatory outside the attribute ranges. Consider a mode choice example, where the decision-maker chooses travel mode based on travel time and cost within the ranges [10, 25] and [2, 4], respectively (omitted units for simplicity). Both attributes follow the half-triangular fuzzy membership function, as illustrated in Figure 1. Since the systematic utility for any travel time value above 25 units would remain the same as the one obtained at 25, we can say that the decision-maker is not making any trade-off between travel time and cost outside the ranges of attributes. Section S.4 in the online appendix discusses other fuzzy membership functions, and the operationalisation of attribute cut-offs in CI is illustrated through an example in Section S.5.

### 2.5 Discussion on the advantages of MNP-CI with attribute cut-offs

In summary, "MNP-CI with attribute cut-offs" is a flexible specification at several levels. The importance attached to the "individual attribute" based on its range is captured by the fuzzy membership function in the normalisation step, flexible interactions between attributes are incorporated by the CI function in the aggregation step, and unrestricted substitution effects are specified using the MNP kernel. The operationalization of attribute cut-offs through the fuzzy membership function makes the MNP-CI model superior to the existing penalty-based semi-compensatory approaches (Swait, 2001; Martínez et al., 2009). For instance, the binomial function in Martínez et al. (2009) induces a non-zero penalty even if the attribute's value is within a certain range. Moreover, this function has an additional parameter (cut-off tolerance to define choice probability at the boundary) to bound the penalty term, but this parameter is assumed fixed to make the estimation stable (Rashedi and Nurul Habib, 2020). In sum, modelling and estimation of one-stage semi-compensatory models with endogenous attribute cut-offs have several challenges, which can be addressed by directly specifying attribute cut-offs through fuzzy membership function in MNP-CI and estimating it through a constrained maximum likelihood estimator.

It is worth noting that irrespective of the type of aggregation function, the attribute cut-off approach can be applied as part of the MNP-WS framework to model the semi-compensatory choice behaviour. However, the incorporation of attribute cut-offs in the weighted sum utility specification leads to numerical issues. Unlike fuzzy measures in CI, the magnitude of parameters in the weighted sum approach could explode because they are unconstrained. The problem is particularly acute in the case of mix of explanatory variables of different nature (continuous, ordinal, and count). These issues are not directly related to the cut-off approach but are a limitation of numerical optimization. Thus, from a practical standpoint, we recommend to use the attribute cut-off approach with a fuzzy-measure-based aggregation function instead of the weighted sum function.

Figure 2 summarises the trade-off between model complexity and interpretability in MNP-WS, MNP-CI with attribute cut-offs and popular machine learning/data-driven/non-parametric algorithms. The CI-based indirect utility function specification with attribute cut-offs offers greater interpretability, but it is slightly more complex than the corresponding linear utility specification due to the involvement of a constrained optimisation problem and the higher number of parameters. However, the proposed specification has lower complexity and higher interpretability relative to data-driven methods because the specification is parametric and transformed parameters have behavioural meaning.

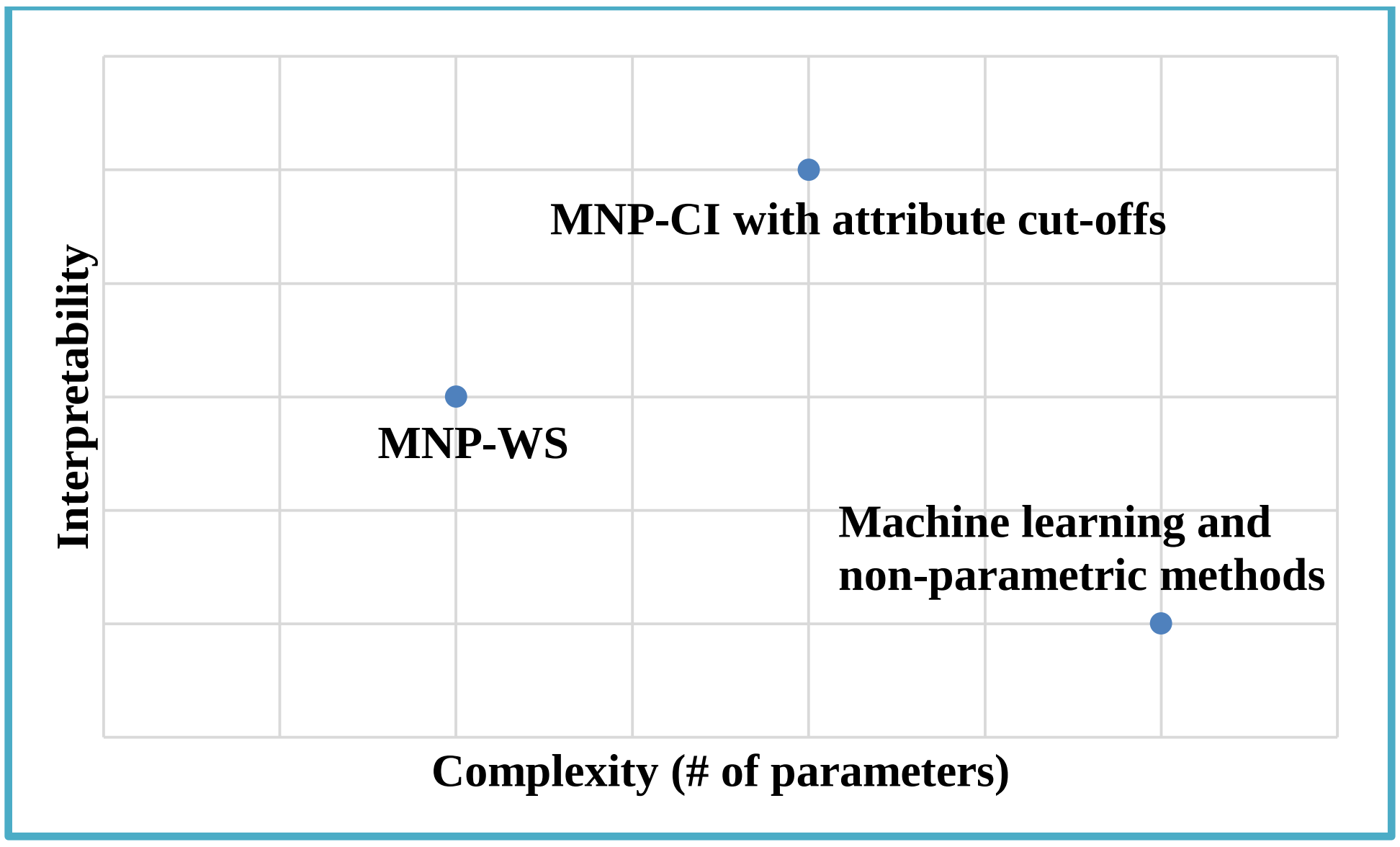

**Figure 2:** Trade-off between complexity and interpretability in different models

# 3. Simulation Study

## 3.1 Simulation Set-up

### 3.1.1 Statistical Measures and Model Specifications

Before adopting MNP-CI models in empirical applications, their statistical properties need to be studied. In this section, we analyse the parameter recovery measured by standard deviation normalized mean absolute error (SDMAE) in a comprehensive Monte Carlo study. SDMAE for a parameter is obtained by dividing the mean of absolute error (mean of absolute difference between true and estimate value across samples) with the standard deviation of true values of all parameters. SDMAE is more appropriate than absolute percentage bias (APB) because APB for low values (i.e. between 0 and 1) of CI parameters tends to over-exaggerate the bias. Other statistics such as symmetric mean absolute percentage error (SMAPE) and root mean square error (RMSE) also suffer from similar issues and lack interpretation (Goodwin and Lawton, 1999; Hyndman and Koehler, 2006; and Armstrong and Collopy, 1992). Further, we also statistically compare the marginal effects based on the true and estimated parameters. For inference evaluation, we compute coverage probability (CP). CP is defined as the proportion of time the confidence interval (95%) contains the true value of parameter. It can be calculated using following expression (Koehler, et. al., 2009):

$$\text{CP} = \frac{1}{R}\sum_{r=1}^{R} I\left[ b_{est} - 1.96 * \text{std.err}(b_{est}) \le b_{true} \le b_{est} + 1.96 * \text{std.err}(b_{est}) \right]$$

where $\beta_{true}$ is the true value of the parameter, $\beta_{est}$ is the estimated value of the parameter, $R$ is the number of samples and the term $\text{std.err}(.)$ indicates the asymptotic standard error (ASE) of the estimated parameter. A CP value close to 0.95 suggests that the estimator is overall reliable in terms of parameter recovery and inference.

We consider four specifications of the MNP-CI model in increasing order of complexity:

1. CI and identical and independent (IID) error structure (CI-IID)
2. CI with cut-off on attributes, and (IID) error structure (CIC-IID)
3. CI with cut-off on attributes, and diagonal error structure (CIC-DE)
4. CI with cut-off on attributes, and full-covariance error structure (CIC-FE)

Comparison of the results of CI-IID and CIC-IID specifications will show the effect of adding attribute cut-offs on statistical properties of model parameters. Step-by-step increment in error complexity from CIC-IID to CIC-FE allows us to comprehend how statistical properties of CI and attribute cut-off parameters are affected due to complexity of error-covariance structure. We also benchmark the performance of each MNP-CI specification against the corresponding MNP-WS specification.

Since the number of CI parameters and constraints quickly increases with the number of attributes due to increase in interaction effects, we analyse the statistical properties of the above-discussed specifications for both four and six attributes. This exercise is imperative to ensure that performance of the model does not deteriorate due to increase in the number of parameters. Table 1 provides the total number of parameters estimated for each of the four model configurations for both attribute cases.

In all four specifications and both attribute configurations, we consider the data generating process (DGP) with five alternatives and sample size of 3000 respondents. For each specification-attribute configuration, MNP-CI and MNP-WS are estimated for 50 datasets. In MNP-WS specifications, interactions are ignored and only the mean effect ($\beta$) parameters are considered. We set the fuzzy-measure value $\mu(.)$ of an attribute in MNP-CI's DGP as its true marginal utility in MNP-WS's DGP, while keeping covariate values same as that of MNP-CI. Although incorporating cut-offs in the MNP-WS model is challenging from the numerical optimization perspective, we did not encounter parameter explosion issues in the Monte Carlo study because all the mean effect parameters in the DGP are between 0 and 1.

Further, all attribute values are drawn from a uniform distribution with a lower and upper limit of 1 and 10 respectively. We restrict the simulation study to MNP-CI with generic fuzzy measures and illustrate the application of alternative-specific MNP-CI in the empirical study. We employ generic CI configuration for two reasons. First, it helps us understand the effect of added complexity (cut-off parameters and error covariance structure) on recoverability of CI parameters. Second, it allows us to test various MNP-CI specifications within a reasonable computational budget.

**Table 1:** Number of parameters and constraints in MNP-CI in the Monte Carlo Study

| # of attributes | Specification | # of CI Parameters | # of Cut-off Parameters | # of Error Parameters | Total # of parameters | # of constraints |
|---|---|---|---|---|---|---|
| 4 | CI-IID | 14 | 0 | 0 | 18 | 32 |
| | CIC-IID | 14 | 12 | 0 | 30 | |
| | CIC-DE | 14 | 12 | 3 | 33 | |
| | CIC-FE | 14 | 12 | 9 | 39 | |
| 6 | CI-IID | 62 | 0 | 0 | 66 | 192 |
| | CIC-IID | 62 | 18 | 0 | 84 | |
| | CIC-DE | 62 | 18 | 3 | 87 | |
| | CIC-FE | 62 | 18 | 9 | 93 | |

**Note:** Total # of parameters includes alternative-specific constants.

### 3.1.2 Data Generating Process

Appendix A.2 provides the data generating process for both four and six attribute cases. We first discuss the details of the simulation configuration for the four-attribute scenario. Since we have five alternatives, we include four alternative-specific intercepts/constants (ASCs) while normalising the first ASC to zero for identification. The four attributes are included in the utility equation through CI in MNP-CI specification. The attribute cut-offs do not appear in the DGP and estimation of CI-IID specification. In the CIC-DE specification, we only estimate the diagonal elements of the error-matrix while fixing non-diagonal elements to 0.5. For the CIC-FE specification, all the elements of the error-matrix are estimated. Note that the first diagonal element of error-covariance matrix is normalised to unity to set the scale of utility in all specifications. For the CI-IID model, the normalisation of attributes is performed using Eq. 3 and for the remaining models, a fuzzy membership function (i.e., half-triangular or trapezoidal) is used depending on the number of cut-off points and the possible sign of the marginal utility of the attribute. In the Monte Carlo study, we also calculate the implied Shapley values and interaction indices using the estimated CI parameters for each dataset to ensure that we do not just assess statistical properties of the CI parameters but also establish the recovery of underlying attribute ranking and complementary effects for pairs of attributes. This exercise is particularly important because it is quite possible that the recovery of CI parameters may be relatively poor due to the high number of parameters, but the statistical properties of MNP-CI will be acceptable for empirical applications if the resulting attribute rankings and complementarity of attribute pairs are recovered well. True Shapley values and interaction indices are also presented along with the DGP in the Appendix A.2.

Next, we provide the details of the simulation configuration for the six-attribute scenario. While the number of alternatives, normalisation strategies and sample size remain the same in both attribute configurations, details about attribute cut-offs is required due to addition of two attributes in six-attribute scenario. For the first four attributes, we use the same membership function as the ones we have for the four-attribute case. For the fifth and sixth attributes, we use half-triangular and trapezoidal membership functions, respectively.

## 3.2 Simulation Results and Discussion

### 3.2.1 Recoverability of Model Parameters

Since the considered MNP-CI specifications involve a large number of parameters, we aggregate statistical measures across group of parameters. Figure 3 reports SDMAE for four groups of parameters for each specification – CI parameters, Shapley value and interaction indices, attribute cut-offs, and error-covariance matrix. A similar plot of APB values is also presented in Figure O.3 of the online appendix (see section S.6). The findings from APB and SDMAE are fairly consistent, but we discuss the latter because the former provides exaggerated values when true parameter values are small (e.g., interaction indices in our case study).

SDMAE for CI parameters (Figure 3a) does not increase substantially with the model complexity for four-attribute scenario, but the recovery of CI parameters for six-attribute case is affected substantially in case of non-IID error covariance structure. As expected, SDMAE of CI parameters is higher for six-attribute scenario as compared to that for four-attribute scenario, simply due to increase in the number of parameters from 14 to 62. Specifically, SDMAE of CI parameters for six-attribute scenario is almost four times higher than that for four-attribute scenario in case of non-IID error structure (i.e., CIC-DE and CIC-FE configurations). There is no specific pattern in SDMAE of MNP-WS and MNP-CI, but we can see that the former has much lower value of SDMAE in the most complex error structure (i.e. CIC-FE configuration) for both attribute scenarios.

On the other hand, recoverability of the Shapley values and interaction indices (Figure 3b) are excellent with an SDMAE below 0.1 for all configurations, except for MNP-CI in CIC-FE configuration. SDMAE is not much sensitive to model complexity and the number of attributes. This result indicates that despite slightly poor recovery of CI parameters for complex error structures and higher number of attributes, the underlying attribute ranking and complementarity of attribute pairs are recovered equally well in the considered scenarios. Such a characteristic is critical to enable the analyst to build a comprehensive MNP-CI model without worrying too much about deterioration in the recovery of Shapley values and interaction indices. It is worth re-emphasizing that CI parameters do not have much behavioural meaning (except the knowledge about the importance of attribute pairs), rather measures like Shapley values and interaction indices are critical from the perspective of policy recommendations.

SDMAE for the attribute cut-off parameters is presented in Figure 3c. First, the recovery of cut-offs for four-attribute scenario is slightly better than that for six-attribute scenario across model configurations for CI-based MNP model. On the other hand, the difference is negligible for MNP-WS model (both across model configurations and the number of attributes). Overall, the recovery of cut-off parameters is excellent irrespective of model complexity. This is a highly encouraging result as it suggests that semi-compensatory behaviour can be recovered in MNP-CI as well as MNP-WS while considering a flexible substitution pattern across alternatives. This result is even more important for the wider applicability of MNP-CI because MNP-WS with attribute cut-offs might encounter numerical issues in the estimation due to the differences in the scale of model parameters.

SDMAE values of error-covariance parameters are similar for MNP-WS and MNP-CI in case of a diagonal error covariance matrix (i.e., CIC-DE configuration); however, the former outperforms the latter in case of full error covariance matrix (i.e., CIC-FE configuration). The recovery of error-covariance parameters in MNP-CI for the most complex configuration is twice as bad as of MNP-WS, suggesting that recovery of full error-covariance matrix is slightly challenging in MNP-CI model.

We also evaluate the difference between true and the estimated marginal effect values for CIC-FE specification and plot them for four- and six-attribute scenarios in Figures 4a and 4b, respectively. In both the scenarios, we change specific attribute by a certain percentage (indicated on the horizontal axis of plots) and evaluate the change in probability of choosing all five alternatives. This process was repeated for all the 50 datasets and the difference between true (computed using true parameter values) and the estimated marginal effect value was evaluated using t-test for each dataset. In order to perform the t-test, for every dataset, the average and standard deviation of the marginal effect values (for both true and estimated) are used as the point estimate ( $\beta_{\text{ME}}$ ) and corresponding standard error ( $\sigma_{\text{ME}}$ ). Then the t-test to check if two estimates are statistically indifferent can be performed using the following expression:

$$\text{t-value} = \frac{abs\left(\beta_{\text{true ME}} - \beta_{\text{estimated ME}}\right)}{\left(\sigma^2_{\text{true ME}} + \sigma^2_{\text{estimated ME}}\right)^{0.5}}$$

If the calculated t-value is smaller than 1.96, we do not have enough statistical evident to reject the null hypothesis at 0.05 significance level that true and estimated marginal effect values are equal. The t-value was obtained for each alternative with respect to each attribute for 50 samples. The t-value is converted into a binary indicator (with 1 representing the inability to reject the null hypothesis and 0 otherwise) and the overall proportion for each case is subsequently obtained. In the case of four attributes, the overall proportion is 0.79 (across all alternatives and attributes), but it is 0.65 for the six-attribute case. This result suggests that the MNP-CI model does a good job for four-attribute case, but the performance slightly deteriorates for the higher number of attributes.

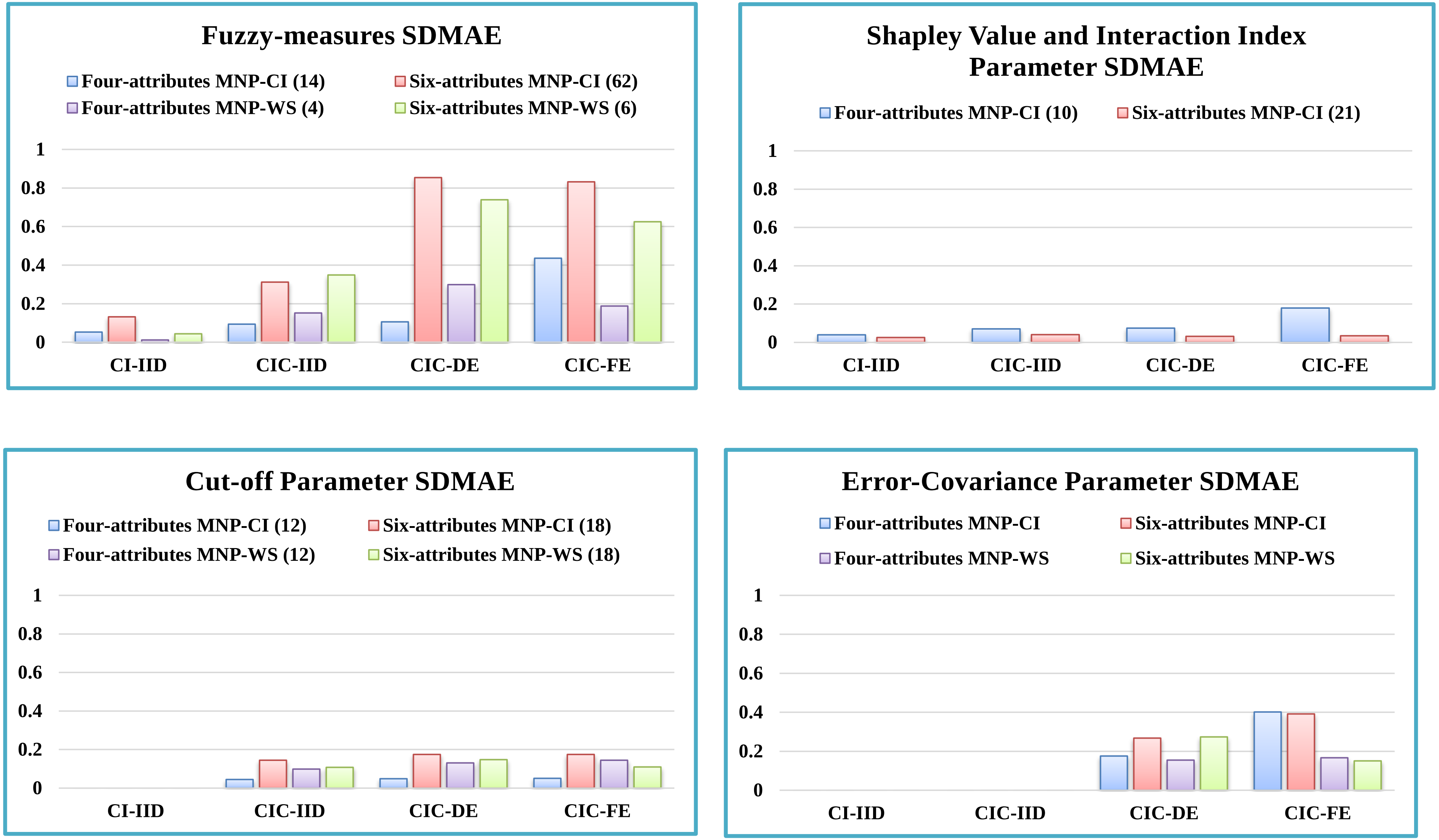

**Figure 3:** Standard deviation normalized mean absolute error (SDMAE) for various parameter groups (the number of parameters in parenthesis)

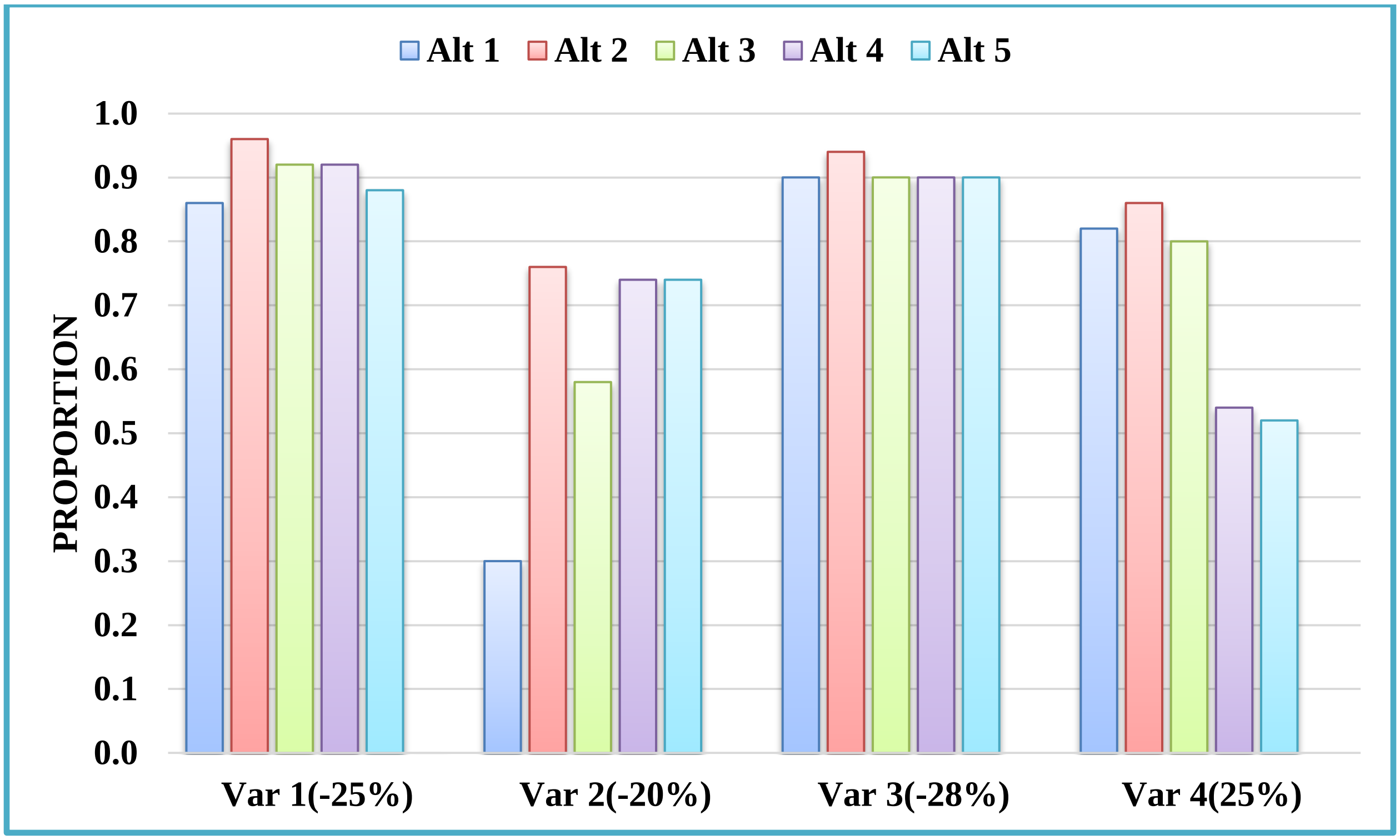

**Figure 4a:** Proportion of samples with statistically insignificant difference between marginal effect values based on true and estimated parameters (four-attribute scenario)

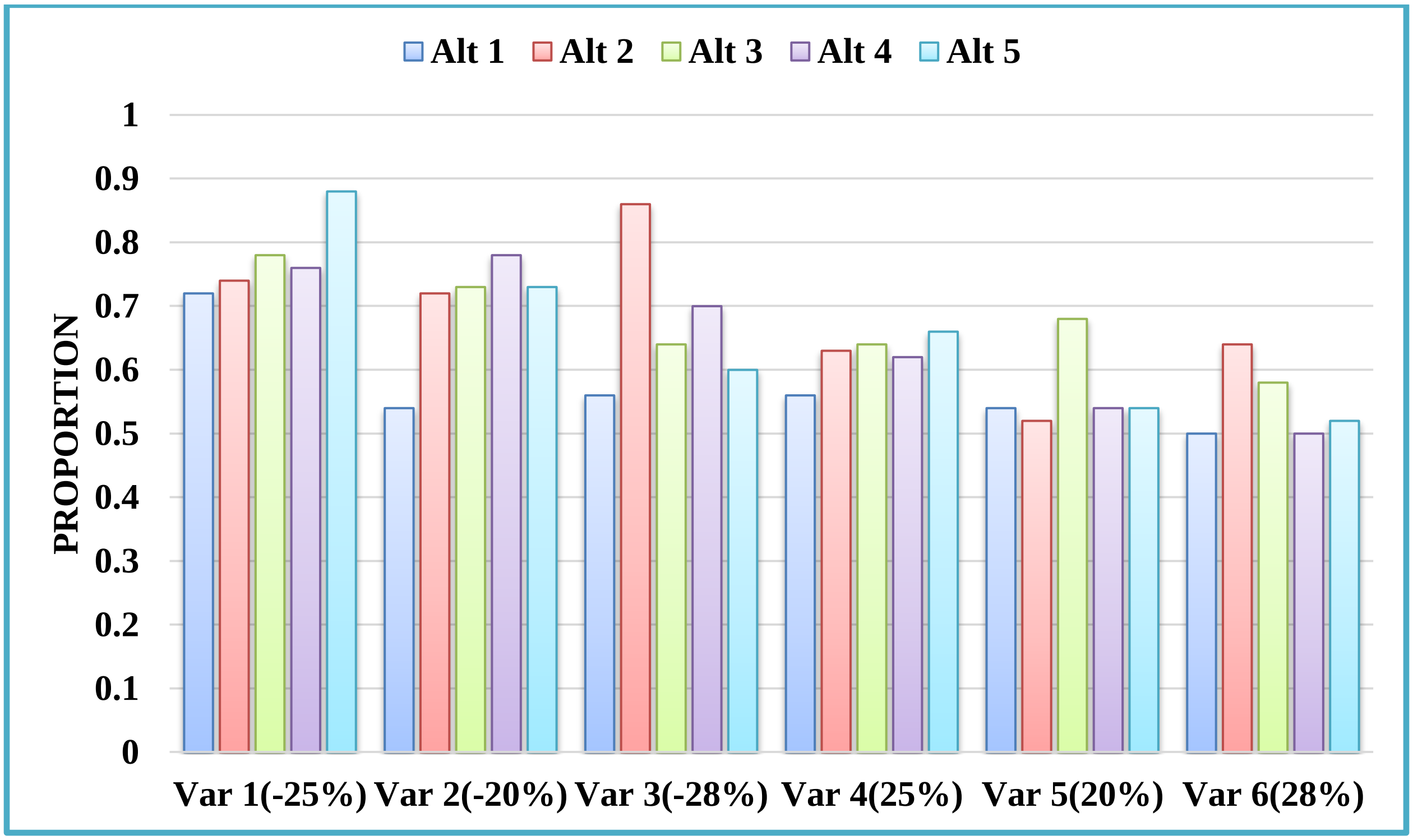

**Figure 4b:** Proportion of samples with statistically insignificant difference between marginal effect values based on true and estimated parameters (six attribute scenario)

### 3.2.2 Coverage Probability

Figure 5 (5a to 5c) shows CP for CI parameters, cut-off parameters and error-covariance parameters. Figure 5a shows that CP values for CI parameters across all specifications of MNP-CI are excellent with a minimum value of more than 0.9. This result reduces worry associated with slightly poor SDMAE values at a few instances. In fact, MNP-CI has better CP values for CI parameters than those of corresponding MNP-WS model specifications. CP values for cut-off parameters in MNP-CI are slightly lower than those of CI parameters – around 0.89 for four-attribute and 0.77 for six-attribute scenario (see Figure 5b). Whereas these CP values are marginally higher for MNP-CI model as compared to MNP-WS model for four-attribute scenario, MNP-WS marginally outperforms MNP-CI for six-attribute scenario (around 0.77 vs. 0.84). This trend can be attributed to more complex interaction effects in six-attribute scenario. Finally, Figure 5c shows that CP for error-covariance parameters is excellent for both MNP-CI and MNP-WS with average CP of 0.98 and 0.90, respectively. This result suggests that relatively larger SDMAE values of MNP-CI for error-covariance parameters in CIC-FE configuration are not concerning.  Overall, the Monte Carlo simulation suggests that statistical properties of MNP-CI estimator are comparable to that of MNP-WS.

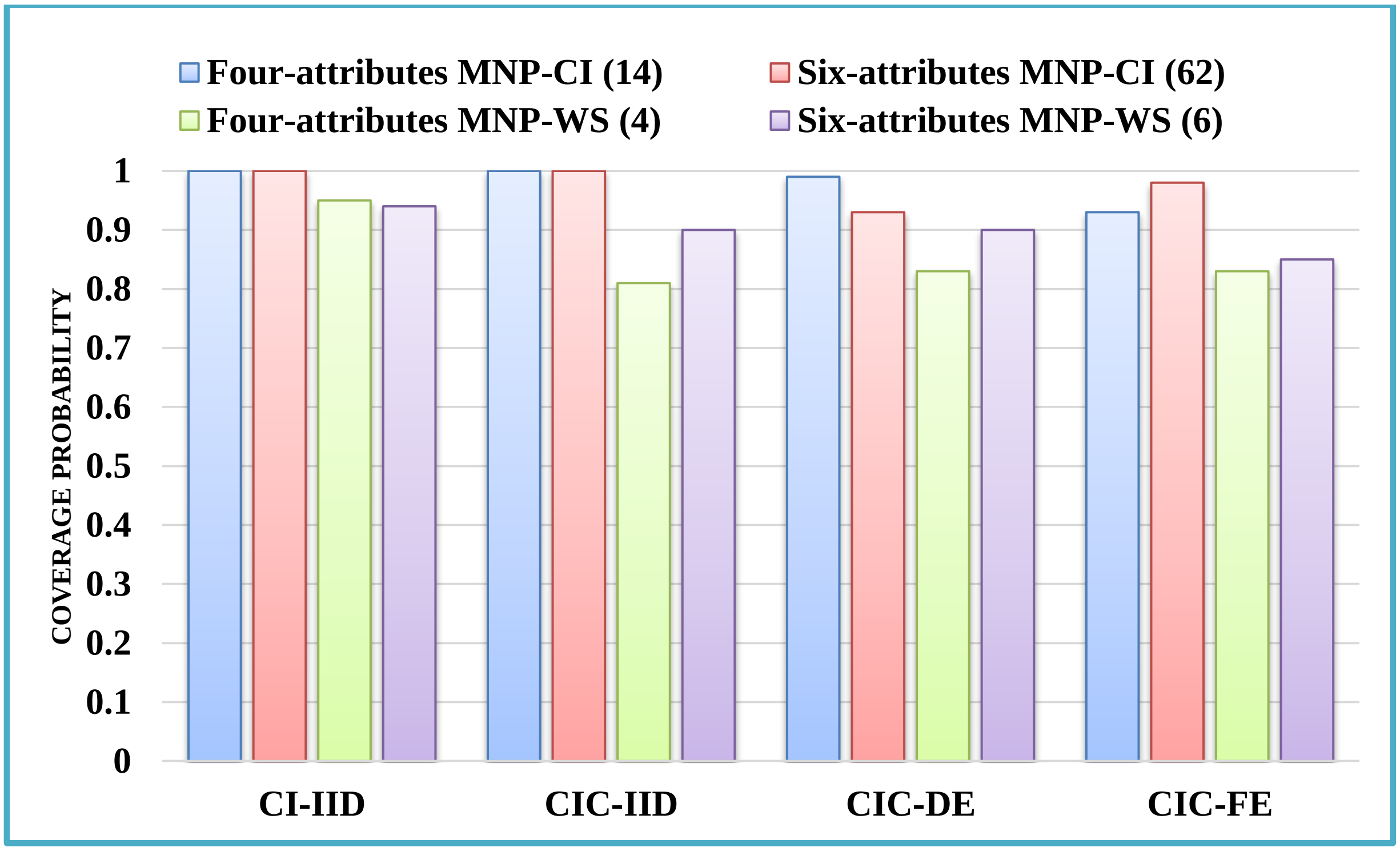

**Figure 5a:** Coverage probability for fuzzy-measures/mean-effect parameters in MNP-CI and MNP-WS models

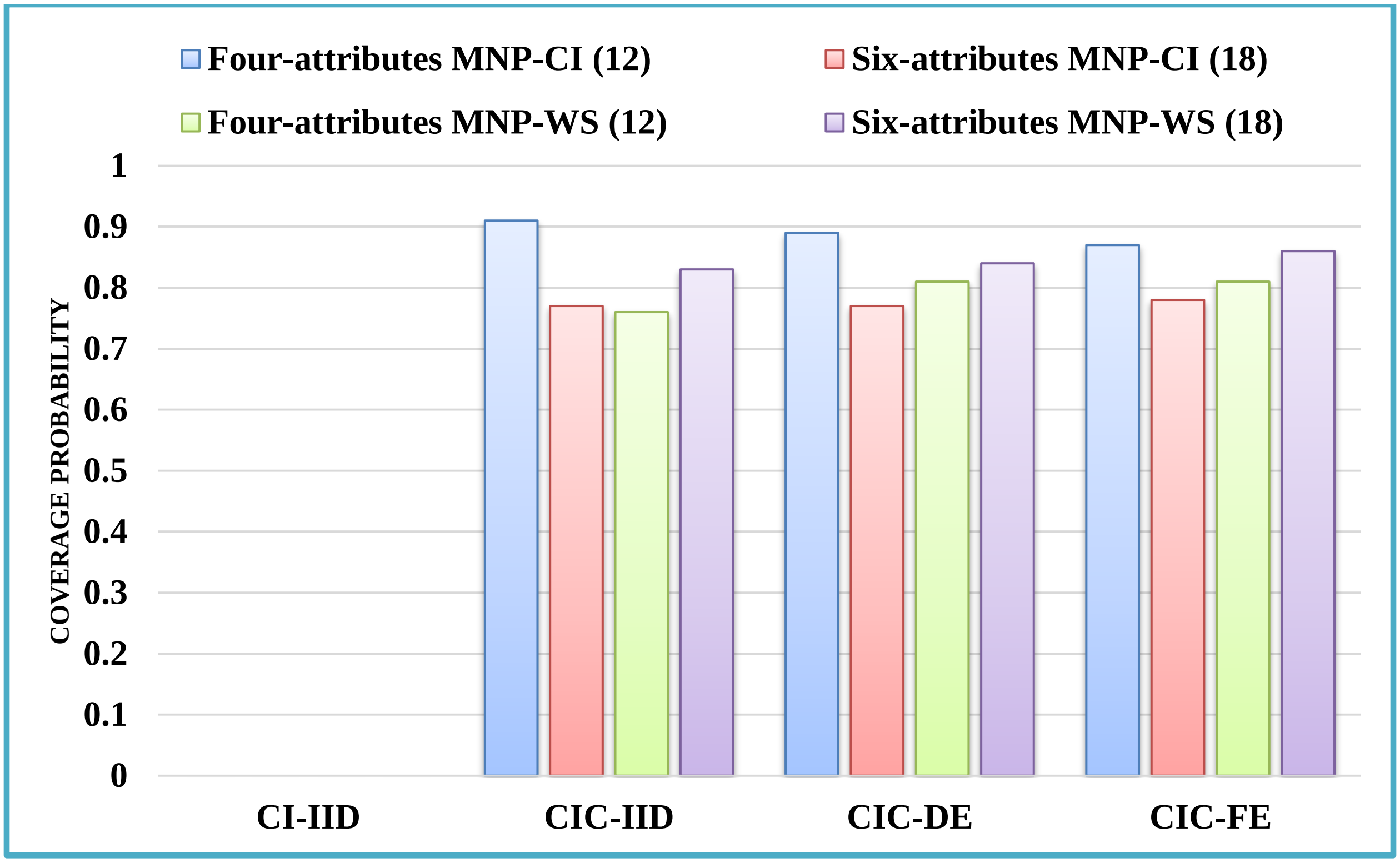

**Figure 5b:** Coverage probability for cut-off parameters in MNP-CI and MNP-WS models

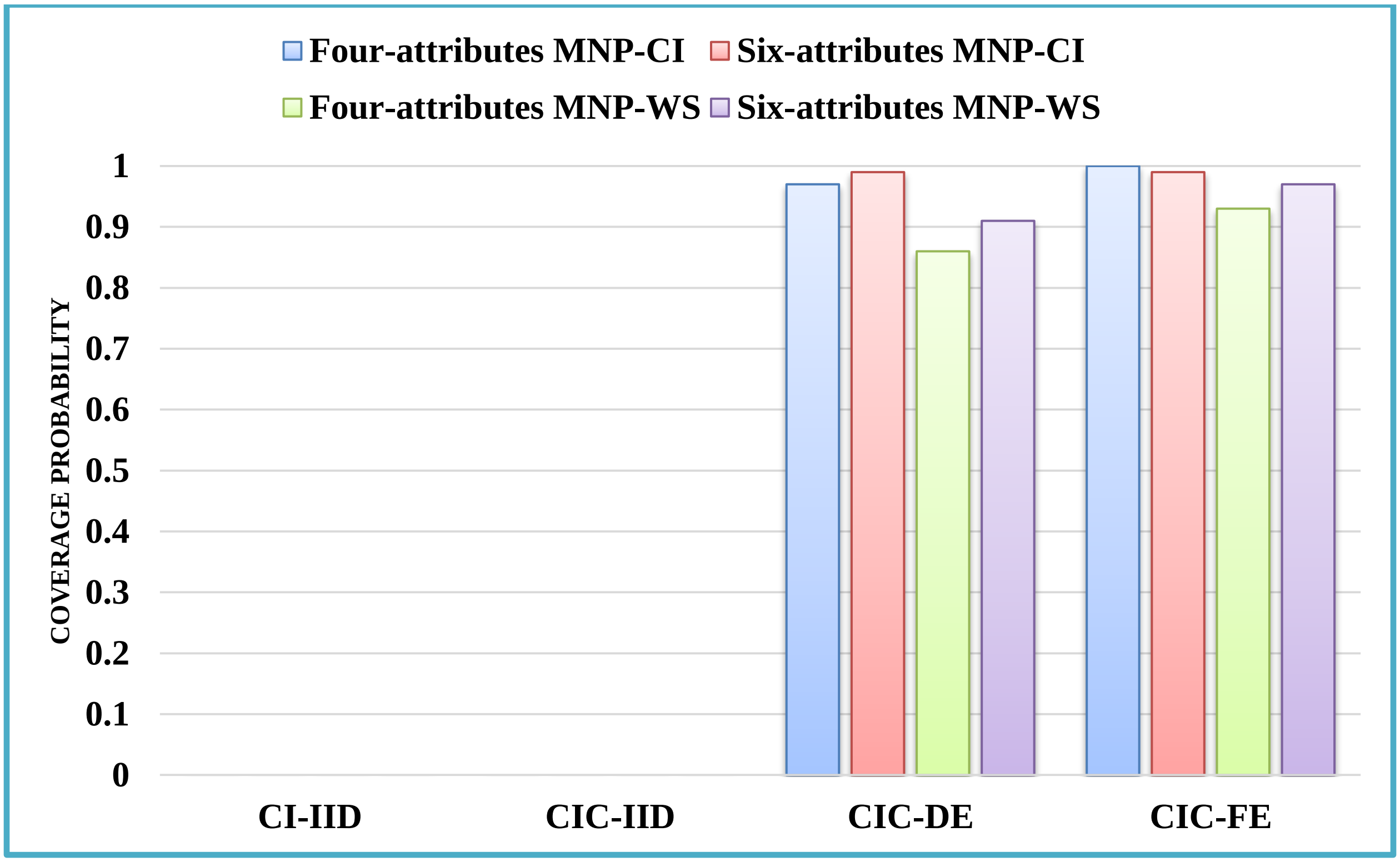

**Figure 5c:** Coverage probability for error-covariance parameters in MNP-CI and MNP-WS models

## 3.3 Assessing the Generality of the MNP-CI Model

So far, we have estimated models that are consistent with the DGP. In this section, we assess the ability of MNP-CI to recover the underlying behavioural process when the true DGP follows MNP-WS specification and vice-versa. Our expectation is that MNP-CI should be able to replicate the weighted attribute aggregation behaviour (MNP-WS) because weighted sum is a special case of CI, but not vice versa. Conditional on the validity of our hypothesis, these simulation results will make a strong case to replace weighted sum utility function with CI in empirical applications.

We consider two simulation studies. In the first study, we generate data using MNP-CI specification and then estimate both MNP-WS and MNP-CI models. In the second study, the DGP follows MNP-WS specification, and both models are estimated. Specifically, we use the four-attribute CIC-FE (with cut-offs and full-error structure) specification in both DGPs as well as estimation. MNP-CI and MNP-WS only differ in terms of the aggregation function. We ignore interaction effects in the MNP-WS estimation and the DGP of the second simulation study (i.e., the systematic utility of MNP-WS has four mean effects). Since MNP-WS and MNP-CI cannot be compared in terms of parameters, we evaluate statistical differences in terms of Akaike information criterion (AIC) values at convergence and the marginal effect values. We perform this comparison for 100 datasets. For marginal effect comparison, we obtain the average change in probability of choosing an alternative due to changing the four attributes by -25%, -20%, -28%, and 25% (one attribute at a time), and conduct a t-test to compare marginal effects of MNP-WS and MNP-CI.

Whereas marginal effects of both simulation studies are provided in Tables O.4 and O.5 (Section S.6) of the online appendix, Table 2 presents the t-test results for both simulation studies. Whereas most t-statistic values are above 1.96 in the first simulation study (when the DGP is MNP-CI), they are below 1.96 in the second simulation study. This result implies that MNP-CI can reproduce the marginal effect values of the MNP-WS model when the DGP follows the latter specification. However, MNP-WS fails to do so when the DGP follows the MNP-CI specification.

**Table 2:** T-statistics for marginal effect difference between MNP-CI and MNP-WS

| Variable | Quantile | When DGP follows MNP-CI | | | | | When DGP follows MNP-WS | | | | |
|---|---|---|---|---|---|---|---|---|---|---|---|
| | | Alt 1 | Alt 2 | Alt 3 | Alt 4 | Alt 5 | Alt 1 | Alt 2 | Alt 3 | Alt 4 | Alt 5 |
| 1 | 0.10 | 4.21 | 4.55 | 3.99 | 3.96 | 4.02 | 0.77 | 1.09 | 0.67 | 0.34 | 0.60 |
| 1 | 0.20 | 4.15 | 4.15 | 3.82 | 3.87 | 3.82 | 0.62 | 0.97 | 0.45 | 0.20 | 0.39 |
| 1 | 0.30 | 3.97 | 3.95 | 3.74 | 3.80 | 3.91 | 0.45 | 0.91 | 0.27 | 0.02 | 0.20 |
| 1 | 0.40 | 3.78 | 3.92 | 3.57 | 3.59 | 3.86 | 0.28 | 0.90 | 0.14 | 0.14 | 0.08 |
| 1 | 0.50 | 3.66 | 3.71 | 3.47 | 3.46 | 3.76 | 0.15 | 0.89 | 0.03 | 0.19 | 0.00 |
| 1 | 0.60 | 3.53 | 3.59 | 3.35 | 3.37 | 3.58 | 0.02 | 0.90 | 0.02 | 0.22 | 0.04 |
| 1 | 0.70 | 3.47 | 3.55 | 3.19 | 3.25 | 3.47 | 0.07 | 0.88 | 0.07 | 0.25 | 0.09 |
| 1 | 0.80 | 3.44 | 3.46 | 3.09 | 3.16 | 3.40 | 0.09 | 0.90 | 0.14 | 0.29 | 0.12 |
| 1 | 0.90 | 3.36 | 3.41 | 3.01 | 3.11 | 3.33 | 0.16 | 0.90 | 0.15 | 0.31 | 0.16 |
| 1 | 0.99 | 3.28 | 3.38 | 2.95 | 3.07 | 3.23 | 0.19 | 0.90 | 0.18 | 0.32 | 0.18 |
| 2 | 0.10 | 3.84 | 4.22 | 3.37 | 3.46 | 3.98 | 0.18 | 0.63 | 0.00 | 0.04 | 0.14 |
| 2 | 0.20 | 3.89 | 4.18 | 3.35 | 3.50 | 3.95 | 0.07 | 0.46 | 0.12 | 0.03 | 0.07 |
| 2 | 0.30 | 3.98 | 4.16 | 3.35 | 3.56 | 3.95 | 0.03 | 0.41 | 0.16 | 0.10 | 0.02 |
| 2 | 0.40 | 3.92 | 4.04 | 3.37 | 3.63 | 3.94 | 0.12 | 0.35 | 0.22 | 0.18 | 0.10 |
| 2 | 0.50 | 3.91 | 4.03 | 3.35 | 3.68 | 3.92 | 0.17 | 0.28 | 0.24 | 0.23 | 0.16 |
| 2 | 0.60 | 3.94 | 4.05 | 3.34 | 3.75 | 3.94 | 0.22 | 0.25 | 0.29 | 0.27 | 0.21 |
| 2 | 0.70 | 3.94 | 4.04 | 3.38 | 3.75 | 3.91 | 0.25 | 0.21 | 0.32 | 0.32 | 0.23 |
| 2 | 0.80 | 3.94 | 4.02 | 3.37 | 3.75 | 3.90 | 0.28 | 0.17 | 0.35 | 0.34 | 0.27 |
| 2 | 0.90 | 3.95 | 3.99 | 3.39 | 3.74 | 3.90 | 0.31 | 0.13 | 0.36 | 0.38 | 0.30 |
| 2 | 0.99 | 3.92 | 3.99 | 3.41 | 3.76 | 3.89 | 0.35 | 0.10 | 0.39 | 0.39 | 0.32 |
| 3 | 0.10 | 5.24 | 5.14 | 4.61 | 4.52 | 4.67 | 0.07 | 0.45 | 0.26 | 0.40 | 0.30 |
| 3 | 0.20 | 4.91 | 4.80 | 4.57 | 4.35 | 4.48 | 0.20 | 0.46 | 0.41 | 0.51 | 0.42 |
| 3 | 0.30 | 4.70 | 4.81 | 4.23 | 4.23 | 4.36 | 0.36 | 0.45 | 0.46 | 0.56 | 0.55 |
| 3 | 0.40 | 4.45 | 4.74 | 4.32 | 4.07 | 4.19 | 0.41 | 0.43 | 0.49 | 0.58 | 0.58 |
| 3 | 0.50 | 4.39 | 4.70 | 4.13 | 3.95 | 4.19 | 0.50 | 0.44 | 0.51 | 0.61 | 0.62 |
| 3 | 0.60 | 4.27 | 4.66 | 4.00 | 3.84 | 4.12 | 0.54 | 0.42 | 0.53 | 0.64 | 0.64 |
| 3 | 0.70 | 4.16 | 4.53 | 3.95 | 3.78 | 3.99 | 0.57 | 0.43 | 0.55 | 0.64 | 0.65 |
| 3 | 0.80 | 4.10 | 4.46 | 3.82 | 3.80 | 3.91 | 0.60 | 0.42 | 0.56 | 0.66 | 0.68 |
| 3 | 0.90 | 4.03 | 4.41 | 3.79 | 3.72 | 3.92 | 0.61 | 0.41 | 0.58 | 0.66 | 0.68 |
| 3 | 0.99 | 3.95 | 4.36 | 3.64 | 3.69 | 3.87 | 0.62 | 0.39 | 0.58 | 0.67 | 0.69 |
| 4 | 0.10 | 4.86 | 5.44 | 4.18 | 3.99 | 4.04 | 0.59 | 0.77 | 0.37 | 0.48 | 0.54 |
| 4 | 0.20 | 4.79 | 5.25 | 4.14 | 4.20 | 3.94 | 0.40 | 0.65 | 0.27 | 0.31 | 0.37 |
| 4 | 0.30 | 4.63 | 5.22 | 4.06 | 4.10 | 3.98 | 0.29 | 0.55 | 0.19 | 0.18 | 0.26 |
| 4 | 0.40 | 4.47 | 5.18 | 3.99 | 4.09 | 4.02 | 0.19 | 0.50 | 0.13 | 0.14 | 0.17 |
| 4 | 0.50 | 4.51 | 5.20 | 4.10 | 4.13 | 4.05 | 0.14 | 0.45 | 0.08 | 0.11 | 0.12 |
| 4 | 0.60 | 4.50 | 5.17 | 4.10 | 4.14 | 4.07 | 0.11 | 0.40 | 0.06 | 0.07 | 0.07 |
| 4 | 0.70 | 4.49 | 5.16 | 4.09 | 4.12 | 4.03 | 0.08 | 0.38 | 0.04 | 0.05 | 0.05 |
| 4 | 0.80 | 4.45 | 5.13 | 4.08 | 4.13 | 4.03 | 0.07 | 0.35 | 0.02 | 0.03 | 0.03 |
| 4 | 0.90 | 4.41 | 5.05 | 4.05 | 4.15 | 4.08 | 0.04 | 0.32 | 0.00 | 0.01 | 0.01 |
| 4 | 0.99 | 4.41 | 5.04 | 4.03 | 4.15 | 4.08 | 0.02 | 0.31 | 0.01 | 0.00 | 0.00 |

The average AIC values for MNP-WS and MNP-CI across 50 datasets are 8991.90 and 8778.74 in the first study, and 8727.79 and 8792.83 in the second study. On average, the MNP-CI model outperforms the MNP-WS model when the true DGP is based on the CI function. Conversely, the difference between MNP-WS and MNP-CI is not substantial when the true DGP is based on the WS function. These findings suggest that MNP-WS may not provide acceptable results when the underlying DGP follows MNP-CI specification, but MNP-CI can recover the underlying weighted-sum DGP.

In sum, Monte Carlo studies establish statistical properties of the MNP-CI model. They also illustrate how MNP-CI can nest MNP-WS and recover flexible substitution patterns and semi-compensatory behaviour through attribute cut-offs. MNP-CI thus has all characteristics to become a workhorse model in discrete choice modelling literature.

# 4. Empirical Application

### 4.1 Data Description

For the empirical application, we use data from travel mode choice experiment conducted by Liu et al. (2019) in October-November 2017. New Yorkers were asked to choose from three travel modes on the most frequent trip in a discrete choice experiment: i) current travel mode (car or public transport), ii) single-occupancy mobility-on-demand (MoD) service (e.g. Uber), and iii) shared MoD service (e.g. Uberpool). Respondents chose an alternative based on six attributes – out-of-vehicle time (OVTT), in-vehicle travel time (IVTT), trip cost, parking cost, powertrain (gas or electric), and automation availability. The final sample included 1507 respondents with each respondent completing seven choice tasks. Readers are referred to Section 2.1 of Liu et al. (2019) for a detailed discussion on attribute level selection and design of the choice experiment.

### 4.2 Results & Discussion

We estimate five MNP-CI specifications with full error-covariance structure – i) generic CI with no attribute cut-offs (CI-NAC), ii) generic CI with constant only attribute cut-offs (CI-CAC), iii) generic CI with parameterised generic attribute cut-offs as a function of respondent's characteristics (CI-GAC), iv) generic CI with parameterised alternative-specific attribute cut-offs as a function of respondent's characteristics (CI-AGAC), and v) alternative-specific CI with parameterised alternative-specific attribute cut-offs as a function of respondent's characteristics (ACI-AGAC). Note that ACI-AGAC is the application of the extended alternative-specific MNP-CI model (as discussed in Section 2.4.1). We introduce flexibility in attribute cut-off representation in a sequential manner to disentangle the contribution of pure CI-based specification and variety of attribute cut-off specifications towards goodness of fit statistics (i.e., loglikelihood at convergence).

The indirect utility of all the estimated models has two components. Whereas the weighted sum component includes alternative-specific constants (ASCs), electric powertrain dummy, and automation dummy, the CI component includes in-vehicle travel time per km (IVTT/Km), out-of-vehicle travel time per km (OVTT/Km) and cost per km (Cost/Km). While combining weighted sum and CI components in the systematic utility, an estimable factor is multiplied with the CI component to adjust for scale differences between two components (Tehrani et al., 2012). The estimated scale factor in our analysis turns out to be statistically indifferent from 1 at a 0.1 significance level, and therefore, set to 1 in all specifications.

The parameter estimates for the weighted sum component are presented in Table 3. By considering current travel mode as the base, ASCs are estimated for both Uber and Uberpool and found to be statistically significant. In all model specifications, the marginal utilities of electric powertrain and automation dummies are negative and statistically significant. Consistent with Liu et al. (2019), these results suggest that New Yorkers have higher preferences for non-electric and non-automated MoD services.

**Table 3:** Parameter estimates in the weighted sum component of the indirect utility (T-statistic in parenthesis)

| Travel Mode | Covariates | CI-NAC | CI-CAC | CI-GAC | CI-AGAC | ACI-AGAC |
|---|---|---|---|---|---|---|
| Uber | Constant | -0.381(-5.8) | -0.440(-10.4) | -0.432(-10.5) | -0.404(-8.3) | -0.406(-7.7) |
| | Electric | -0.045(-1.3) | -0.054(-1.7) | -0.044(-1.4) | -0.063(-2.0) | -0.057(-1.8) |
| | Automated | -0.117(-3.4) | -0.099(-3.0) | -0.089(-2.7) | -0.113(-3.4) | -0.110(-3.2) |
| Uberpool | Constant | -0.621(-8.0) | -0.424(-9.9) | -0.572(-11.7) | -0.487(-9.5) | -0.516(-8.9) |
| | Electric | -0.068(-2.4) | -0.029(-1.7) | -0.042(-1.9) | -0.027(-1.2) | -0.033(-1.5) |
| | Automated | -0.011(-0.3) | -0.049(-2.6) | -0.064(-2.6) | -0.058(-2.4) | -0.052(-2.2) |

**Note:** current travel model is base.

Further, in all the estimated models, we consider full error-covariance structure, with a traditional MNP identification strategy – only the difference of error-covariance matrix is estimable, and the top-left element is normalised to 1 (see Appendix A.1 for a detailed discussion). The estimated differenced (and normalised) error covariance matrices are presented in Table 4. The results indicate that errors are correlated[5], and, thus accounting for flexible substitution patterns is crucial to obtain correct elasticity estimates.

**Table 4:** The estimated differenced error-covariance for various models (t-statistics in parenthesis)

| Model | Error-covariance |
|---|---|
| Generic CI with no attribute cut-offs (**CI-NAC**) | $\begin{bmatrix} 1.00 \text{ (fixed)} & \\ -0.041\ (-1.28) & 0.697\ (6.69) \end{bmatrix}$ |
| Generic CI with generic constant only attribute cut-offs (**CI-CAC**) | $\begin{bmatrix} 1.00 \text{ (fixed)} & \\ -0.406\ (-4.50) & 0.284\ (5.40) \end{bmatrix}$ |
| Generic CI with generic demographics-based attribute cut-offs (**CI-GAC**) | $\begin{bmatrix} 1.00 \text{ (fixed)} & \\ -0.602\ (-7.61) & 0.513\ (6.29) \end{bmatrix}$ |
| Generic CI with alternative-specific demographics-based attribute cut-offs (**CI-AGAC**) | $\begin{bmatrix} 1.00 \text{ (fixed)} & \\ -0.572\ (-7.81) & 0.479\ (6.05) \end{bmatrix}$ |
| Alternative-specific CI with alternative-specific demographics-based cut-offs (**ACI-AGAC**) | $\begin{bmatrix} 1.00 \text{ (fixed)} & \\ -0.590\ (-7.40) & 0.489\ (6.01) \end{bmatrix}$ |

*4.2.1 Shapley Values and Interaction Indices*

We focus on three variables that are included in the CI component of the utility. We do not discuss the estimated fuzzy measures here due to lack of interpretability (but are available in Table O.6 in Section S.7 of the online appendix). Instead, the resulting Shapley values and interaction indices are presented in Table 5, and are discussed in detail. Three interesting trends can be observed in the Shapley values across different model specifications. First, Cost/Km is the most important variable (from the respondent's point of view), followed by OVTT/Km and IVTT/Km. This result is consistent with the literature, as cost and OVTT are reported to be the two main determinants of mode choice (Gang, 2007; Xie et al., 2019; Dong, 2020). Shapley values are empirically advantageous to determine such rankings directly, without having to calculate the marginal effect values. Second, as we start to increase the degrees of freedom in MNP-CI through attribute cut-offs, we observe that the attribute ranking remains the same, but

[5] The diagonal and off-diagonal elements of differenced independent and identically distributed (IID) error-structure covariance are 1 and 0.5, respectively. This matrix is equivalent to an identity matrix in un-differenced form.

IVTT/Km becomes less important (i.e., lower Shapley value) in explaining an individual's choice. Third, results of the alternative-specific MNP-CI specification (ACI-AGAC) suggest that the distance between OVTT/Km and IVTT/Km is more pronounced for the current travel mode than Uber and Uberpool. Close Shapley values indicate that OVTT and IVTT play a similar role in determining travellers' preferences for MoD services.

The interaction indices also exhibit two interesting trends. First, in the absence of attribute cut-offs, all the pairs have a positive interaction index. Second, as we introduce attribute cut-offs, only OVTT/Km and Cost/Km exhibit a significant complementarity effect. Thus, comparison of both Shapley values and interaction indices between CI-NAC and specifications with attribute cut-offs clearly indicates the importance of accounting for the semi-compensatory behaviour, taste heterogeneity, and alternative-specific effects to correctly identify the importance attached to attributes by the decision-maker in the decision process.

**Table 5:** Shapley values and interaction indices in the empirical study

| **Variables** | **CI-NAC** | **CI-CAC** | **CI-GAC** | **CI-AGAC** | **ACI-AGAC** | |
|---|---|---|---|---|---|---|
| | **All modes** | **All modes** | **All modes** | **All modes** | **Current mode** | **Uber and Uberpool** |
| | **Shapley values** | | | | | |
| IVTT/Km | 0.207 | 0.117 | 0.114 | 0.125 | 0.130 | 0.206 |
| OVTT/Km | 0.284 | 0.319 | 0.282 | 0.336 | 0.347 | 0.265 |
| Cost/Km | 0.509 | 0.564 | 0.604 | 0.539 | 0.523 | 0.529 |
| | **Interaction Indices** | | | | | |
| IVTT/Km, OVTT/Km | 0.182 | -0.034 | -0.029 | 0.093 | 0.122 | 0.046 |
| IVTT/Km, Cost/Km | 0.124 | 0.106 | 0.045 | -0.093 | -0.122 | 0.033 |
| OVTT/Km, Cost/Km | 0.068 | 0.223 | 0.228 | 0.328 | 0.306 | 0.270 |

### *4.2.2 Attribute Cut-off Values*

Next, we turn our attention to attribute cut-off values. Since all the three variables are likely to cause disutility with an increase in their values, we employ two-point half-triangular cut-offs for all three variables (see Figure 1 in Section 2.4.2). First, in the constant only attribute cut-off model (i.e. CI-CAC), the lower and upper thresholds for IVTT/Km are 2.17 and 5.75. This result suggests that the lowest and the highest disutility induced by IVTT/km can be computed by plugging 2.17 and 5.75 values for IVTT/km in CI. Specifically, the normalised IVTT/km value becomes zero when the true IVTT/km is 5.75, and thus, its contribution to CI is 0 (i.e., maximum disutility) for all IVTT/km values above 5.75. The thresholds for OVTT/Km [1.29, 5.71] and Cost/Km [0.07, 2.15] can be interpreted in a similar fashion.

When we parameterise the attribute cut-offs as a function of socio-demographic variables in CI-GAC specification, we obtain several interesting findings regarding the heterogeneity in thresholds on Cost/km (see Table 6). First, households with annual income below 125 thousand dollars have smaller upper threshold for Cost/Km as compared to higher-income households, everything else being constant. Second, older males have lower upper threshold on Cost/Km as compared to younger females, when controlling for income and distance to transit stops. Considering that the lower threshold is the same for both demographic groups, this result implies that the marginal effect of Cost/Km is much higher for older males compared to that for younger females.

Similarly, we observe several interesting relations between socio-demographics and OVTT/Km thresholds. First, households living within 0.5 km radius of the bus stop or subway have a higher threshold for OVTT/Km. Second, males tend to be slightly less patient than females when it comes to OVTT/Km. This result is consistent with the findings of Dittrich and Leopold (2014). Finally, people tend to get impatient with walking and waiting time as they get older, possibly because younger people can better utilize OVTT via mobile phones and tablets.

Table 7 shows the sampling distribution of attribute cut-off values for the CI-GAC specification, which are obtained by transforming the estimates reported in Table 6. It is worth noting that the lower cut-off is always kept constant and the upper cut-off is parameterised by demographics to ensure that the upper cut-off is always greater than the lower cut-off. Since, we did not find any statistically significant cut-off heterogeneity for IVTT/Km, its cut-off values are kept constant across respondents. While there is a substantial variation in thresholds for OVTT/Km and Cost/Km in CI-GAC specification, median values are close to the one obtained by CI-CAC specification.

Finally, we allow the attribute cut-off values to be alternative-specific in the CI-AGAC and ACI-AGAC specification and capture heterogeneity in alternative-specific cut-offs across different socio-demographic groups. We do not discuss the heterogeneity results of CI-AGAC and ACI-AGAC in detail here as they are intuitive and are consistent with those of CI-GAC specification (i.e., the one with generic attribute cut-offs). They are available in Table O.7 to O.12 in Section S.7 of the online appendix. We note that preference ranges for attributes vary substantially across alternatives (see Table O.13 to O.18 in Section S.7 of the online appendix). For example, in contrast to generic lower and upper thresholds of {1.33, 3.96} for IVTT/Km in CI-GAC specification, these values for current travel mode, Uber and Uberpool are {3.03, 3.53}, {0.85, 5.19}, and {0.52, 8.37} in ACI-AGAC specification, respectively.

**Table 6:** Attribute cut-off heterogeneity in CI-GAC model (T-statistic in parenthesis)

| **Explanatory variables** | | **IVTT/Km** | | **OVTT/Km** | | **Cost/Km** | |
|---|---|---|---|---|---|---|---|
| | | **Lower Cut-off** | **Upper Cut-off** | **Lower Cut-off** | **Upper Cut-off** | **Lower Cut-off** | **Upper Cut-off** |
| | Constant | 0.29 (1.0) | 0.97 (3.9) | 0.44 (10.8) | 1.77 (16.5) | -1.73 (-6.1) | 1.2 (24) |
| Household income * travel distance (in USD * km) [Base: >125K] | (<=50K) * distance | | | | | | -0.28 (-4.9) |
| | (> 50K & <=125K) * distance | | | | | | -0.25 (-4.8) |
| Distance to Bus stop (in km) [Base: ≤ 0.5] | (> 0.5 & ≤ 1) | | | | -1.42 (-5.7) | | |
| | (> 1 & ≤ 2) | | | | -0.35 (-1.5) | | |
| | (> 2) | | | | -0.35 (-1.5) | | |
| Distance to subway (in km) [Base: ≤ 0.5] | (> 0.5 & ≤ 1) | | | | -0.3 (-2.1) | | |
| | (> 1 & ≤ 2) | | | | -1.97 (-5.2) | | |
| | (> 2) | | | | -1.17 (-5.3) | | |
| Male | | | | | -0.23 (-1.9) | | -0.16 (-2.5) |
| Years since owing a driver's license | | | | | 0.01 (1.5) | | |
| Age (in years) [Base: 23 – 38] | Age (7 - 22) | | | | 1.09 (5.4) | | |
| | Age (39 - 54) | | | | | | -0.56 (-8.2) |
| | Age (55 - 73) | | | | -0.6 (-2.0) | | -0.98 (-9.6) |

**Table 7:** Distribution of attribute cut-off for generic CI with demographics-based attribute cut-offs (CI-GAC model)

| Percentile | IVTT/Km | | OVTT/Km | | Cost/Km | |
|---|---|---|---|---|---|---|
| | Lower Cut-off | Upper Cut-off | Lower Cut-off | Upper Cut-off | Lower Cut-off | Upper Cut-off |
| 10 | 1.33 | 3.96 | 1.55 | 2.44 | 0.18 | 1.24 |
| 20 | 1.33 | 3.96 | 1.55 | 2.94 | 0.18 | 1.57 |
| 30 | 1.33 | 3.96 | 1.55 | 4.07 | 0.18 | 1.80 |
| 40 | 1.33 | 3.96 | 1.55 | 5.68 | 0.18 | 2.08 |
| 50 | 1.33 | 3.96 | 1.55 | 6.31 | 0.18 | 2.51 |
| 60 | 1.33 | 3.96 | 1.55 | 7.13 | 0.18 | 2.77 |
| 70 | 1.33 | 3.96 | 1.55 | 7.44 | 0.18 | 2.99 |
| 80 | 1.33 | 3.96 | 1.55 | 8.10 | 0.18 | 3.11 |
| 90 | 1.33 | 3.96 | 1.55 | 9.06 | 0.18 | 3.34 |
| 95 | 1.33 | 3.96 | 1.55 | 11.44 | 0.18 | 3.50 |
| 99 | 1.33 | 3.96 | 1.55 | 11.85 | 0.18 | 3.50 |
| 100 | 1.33 | 3.96 | 1.55 | 12.47 | 0.18 | 3.50 |

### *4.2.3 Marginal Effects*

We compare the marginal effects (change of probability) obtained by the CI-NAC and the ACI-AGAC specifications for scenarios where one aspect of the MoD service quality is improved at a time. Table 8 provides the sampling distribution of change in probability of choosing all three options when the IVTT/Km is reduced by 10% for both Uber and Uberpool. The change in probability implied by the CI-NAC specification is higher in magnitude as compared to the ACI-AGAC model at almost every percentile point. This trend is in line with expectations because attribute cut-offs moderate the change in value of the variable through preference ranges depending upon the demographics. Thus, the difference between the policy implications of CI-NAC and ACI-AGAC specifications is considerable.

**Table 8:** Change in probability due to 10% decrease in IVTT/Km of Uber and Uberpool

| Percentile | Current Mode | | Uber | | Uberpool | |
|---|---|---|---|---|---|---|
| | CI-NAC | ACI-AGAC | CI-NAC | ACI-AGAC | CI-NAC | ACI-AGAC |
| 10 | -0.0141 | -0.0119 | 0 | 0 | 0 | 0 |
| 20 | -0.0103 | -0.0064 | 0 | 0 | 0 | 0 |
| 30 | -0.0084 | -0.0011 | 0 | 0 | 0 | 0 |
| 40 | -0.0073 | 0 | 0 | 0 | 0 | 0 |
| 50 | -0.0052 | 0 | 0 | 0 | 0 | 0 |
| 60 | -0.0035 | 0 | 0 | 0 | 0 | 0 |
| 70 | 0 | 0 | 0.0042 | 0 | 0.0041 | 0 |
| 80 | 0 | 0 | 0.0095 | 0 | 0.0075 | 0.0049 |
| 90 | 0 | 0 | 0.0131 | 0 | 0.0143 | 0.0089 |
| 95 | 0 | 0 | 0.0193 | 0.0137 | 0.0183 | 0.0123 |
| 99 | 0 | 0 | 0.0318 | 0.0362 | 0.0380 | 0.0197 |
| 100 | 0 | 0 | 0.0608 | 0.1675 | 0.0656 | 0.0636 |

### *4.2.4 Goodness-of-fit Statistics*

We also evaluate whether increasing the flexibility in attribute cut-off specification translates into better goodness of fit by providing the log-likelihood and Akaike information criterion (AIC) values for all models in Table 9. The results show that the ACI-AGAC specification has the lowest AIC value, demonstrating the significance of considering alternative-specific preference heterogeneity (parameterised attribute cut-offs) and alternative-specific importance of attributes (Choquet parameters). However, the difference between CI-AGAC and ACI-AGAC model is marginal (3 point difference in AIC statistics). It suggests that for the current dataset, alternative-specific preference heterogeneity (i.e., attribute cut-offs) is much more dominant in explaining the choices than alternative-specific attribute ranking (i.e., CI). Another interesting observation is that the AIC value of the CI-CAC model is substantially higher than that of the CI-NAC model (difference of 44 points). It indicates that imposing a constant preference range for all respondents leads to worse goodness-of-fit than the one with no attribute cut-offs. Further, to benchmark the MNP-CI model against traditional MNP-WS model, we also estimate two MNP-WS specification with and without interactions between attributes (see Table O.19 in Section S.7 of the online appendix for the results of MNP-WS). Both MNP-WS specifications have lower loglikelihood and higher AIC values than all considered MNP-CI specifications. It is worth acknowledging that the MNP-WS model with attribute cut-offs is not presented because its estimation encountered numerical issues.

**Table 9:** Comparison of goodness-of-fit statistics across model

| Model | Log-likelihood at convergence | Number of parameters | AIC |
|---|---|---|---|
| Generic CI with no attribute cut-offs (**CI-NAC**) | -8736.36 | 15 | 17502.72 |
| Generic CI with generic constant only attribute cut-offs (**CI-CAC**) | -8752.09 | 21 | 17546.18 |
| Generic CI with generic demographics-based attribute cut-offs (**CI-GAC**) | -8662.06 | 34 | 17392.12 |
| Generic CI with alternative-specific demographics-based attribute cut-offs (**CI-AGAC**) | -8521.05 | 49 | 17140.10 |
| Alternative-specific CI with alternative- specific demographics-based cut-offs (**ACI-AGAC**) | -8511.71 | 57 | 17137.42 |
| **MNP-WS** model with **no interactions**, non-IID error structure and no cut-offs | -8995.90 | 11 | 18013.80 |
| **MNP-WS** model with **all interactions**, non-IID error structure and no cut-offs | -8812.60 | 15 | 17655.20 |

## 5. Conclusion and future works

We present an extension of the multinomial probit (MNP) model where the systematic part of the indirect utility is modelled using the Choquet integral (CI). CI is an appropriate aggregation function as it can flexibly capture attribute interactions while ensuring monotonicity in attribute values and the number of attributes. We further advance the MNP-CI model to account for semi-compensatory behaviour by specifying individual-specific attribute cut-offs through fuzzy membership functions parameterised by demographics. Thus, the proposed MNP-CI model with attribute cut-offs can simultaneously capture: (i) attribute-level evaluation by the decision-maker and heterogeneity in evaluation across socio-demographic groups, (ii) theory-driven flexible aggregation of attributes in the systematic utility, and (iii) unrestricted substitution patterns.

We estimate the proposed model using a constrained maximum likelihood estimator. A comprehensive Monte Carlo study is performed to establish the statistical properties of the estimator. In another simulation study, we demonstrate the generality of the MNP-CI model by showing that it nests the traditional MNP model with weighted-sum utility. The empirical advantages of the proposed model are illustrated in a travel mode choice study that focuses on understanding the preferences of New Yorkers to shift from the current travel model to on-demand mobility.

The MNP-CI model with attribute cut-offs offers several insights. First, the analyst can elicit semi-compensatory choice behaviour using datasets from traditional choice experiments. Second, the estimation of MNP-CI provides the Shapley values of attributes, which translate into attribute importance ranking. Moreover, interaction indices are also a by-product of the MNP-CI estimation that help in identifying whether simultaneous information on a set of attributes is more meaningful for the decision-maker in making a choice. The complementarity between pairs of attributes coupled with their individual importance ranking can help policymakers make informed decisions to improve the preference level of an alternative. In sum, we make a convincing case for the MNP-CI model with attribute cut-offs to become a workhorse model in the discrete choice analysis. The generality and monotonicity of the CI function make the case stronger. This work will spark the interest of researchers to explore other fuzzy-integral-based aggregation functions, such as CI with bipolar scale (i.e., fuzzy measures range between -1 and 1) (Grabisch and Labreuche, 2005).

We discuss four main limitations or challenges of the proposed MNP-CI model (and the related Fuzzy-measures), which also open up avenues for future research. First, the traditional CI can only handle continuous attributes, but incorporating other attribute types, such as ordinal, categorical and count, is non-trivial. To address this challenge, we specify the observed part of the utility function as a combination of weighted sum and CI function in our empirical study, where the continuous attributes are used in the CI function, and other attributes are used in the weighted sum function. Although this approach is practical, advancements in fuzzy integrals can be explored to automatically learn interactions between non-continuous attributes. For instance, Wang et al. (2006) proposed an approach to convert a mix of non-continuous variables (depending upon their observed scale or count value) into a number between 0 and 1 using a fuzzy logic alpha-cut approach. Such advanced methods to directly incorporate non-continuous variables in CI can be explored in future studies.

Second, we could capture systematic heterogeneity in attribute cut-offs and derive behavioural insights by parameterising cut-offs, but capturing unobserved taste heterogeneity through random parameters is not straightforward in MNP-CI. The non-additivity and constrained range (between 0 and 1) of the fuzzy measures make the use of random parameters challenging. Future research may introduce random heterogeneity into the attribute cut-off function, but this additional flexibility comes at the expense of high computation time because the estimation of the extended model will require another layer of simulation.

Third, the analyst needs to pre-determine the sign of the marginal utility of an attribute to apply the fuzzy membership function. This constraint could be challenging for studies with relatively new explanatory variables (e.g., new technologies such as automation). However, we think that CI with a bipolar scale could provide insights into selecting the type of membership function. In particular, the analyst can first estimate the MNP model with bipolar CI and no attribute cut-off to identify the sign of fuzzy measures. Subsequently, attributes can be assigned a fuzzy membership function according to the direction of fuzzy measures in the first step.

Fourth, we have illustrated that the proposed MNP-CI model with attribute cut-offs has good statistical properties for four and six attributes. However, these properties might deteriorate for a large number of attributes due to a steep increase in the number of parameters and constraints. Whereas most empirical studies have six or fewer attributes, and there is a flexibility to include more control variables in the weighted-sum component, future studies need to explore scalable fuzzy measures to ensure their broader and seamless applicability in discrete choice analysis. Readers are referred to recent advancements by Beliakov and Wu (2019) and Beliakov and Divakov (2020), who propose new methods to control the rate of increase in the number of parameters and constraints with the number of attributes.

## Acknowledgement

This research was supported by the CriticalMaaS project (grant number 804469), which is financed by the European Research Council and the Amsterdam Institute of Advanced Metropolitan Solutions.

# Appendix

## A.1 Generalized Multinomial Probit Model with Choquet Integral

### A.1.1 Model Formulation

If $i$ be the index for alternative $i \in \{1, 2, ..., I\}$ and $g$ be the index for attributes $g \in \{1, 2, ..., G\}$, individual $n \in \{1, 2, ..., N\}$ derives the following indirect utility by choosing $i^{th}$ alternative (suppresed individual-level subscript for notational simplicity):

$$U_i = CI_i + \varepsilon_i \tag{A.1}$$

$$CI_i = \sum_{g=1}^{G} h\left(x^i_{\pi_{N_g}}\right)\left(\mu_i\left(A^i_g\right) - \mu_i\left(A^i_{g-1}\right)\right) \tag{A.2}$$

$$h\left(x^i_{\pi_{N_g}}\right) \to 0 \le \left[h\left(x^i_{\pi_{N_1}}\right) \ge h\left(x^i_{\pi_{N_2}}\right) \ge ... \ge h\left(x^i_{\pi_{N_G}}\right)\right] \le 1$$

$$A^i_G = \left\{x^i_{N_1}, x^i_{N_2}, ..., x^i_{N_G}\right\}$$

where the function $h(.)$ is applied on the normalised attribute values to arrange them in decreasing order, $A^i_g$ is the set of attributes of cardinality $g$ for the $i^{th}$ alternative, $\mu_i\left(A^i_g\right)$ is the corresponding fuzzy measure and $\varepsilon_i$ is a normally-distributed error term.

In Eq. A.2, the function $h\left(x^i_{\pi_{N_g}}\right)$ is bounded between 0 and 1. Thus, before calculating the CI value, one needs to rescale the attribute values. Below, we define a set of notations to simultaneously rescale the attribute values using fuzzy membership function and re-write Eq. A.1 in matrix form. We define the following vector/matrix notations:

$\boldsymbol{U} = (U_1, U_2, ..., U_I)$ $[(I \times 1)\text{ vector}]$ ,

$\boldsymbol{x}_i = (x^i_1, x^i_2, ..., x^i_G)[(1 \times G)\text{ vector}]$ , $\quad \boldsymbol{x} = (\boldsymbol{x}_1, \boldsymbol{x}_2, ..., \boldsymbol{x}_I)[(I \times G)\text{ matrix}]$ ,

$\boldsymbol{\mu}_i = (\mu_1, \mu_2, ..., \mu_{2^G-1})\left[1 \times \left(2^G - 1\right)\text{ vector}\right]$, $\quad \boldsymbol{\mu} = (\boldsymbol{\mu}_1, \boldsymbol{\mu}_2, ..., \boldsymbol{\mu}_I,)\left[I \times \left(2^G - 1\right)\text{ matrix}\right]$,

$\boldsymbol{\varepsilon} = (\varepsilon_1, \varepsilon_2, ..., \varepsilon_I)[(I \times 1)\text{ vector}]$ .

Here for notational simplicity, we assume that all alternatives have $G$ attributes but that can be relaxed in alternative-specific CI. Further, we assume a trapezoidal membership function for all the explanatory variables, i.e. four- thresholds per explanatory variable ( $g$ ) for alternative $i$ (i.e., $\psi_{g_i,1} \le \psi_{g_i,2} \le \psi_{g_i,3} \le \psi_{g_i,4}$). Now, stack the threshold elements as follow:

$$\boldsymbol{\psi}_{g_i} = \left(\psi_{g_i,1}, \psi_{g_i,2}, \psi_{g_i,3}, \psi_{g_i,4}\right)\left[(1\times 4) \text{ vector}\right], \boldsymbol{\psi}_i = \left(\boldsymbol{\psi}_{1_i}, \boldsymbol{\psi}_{2_i}, ..., \boldsymbol{\psi}_{G_i}\right)\left[(G\times 4) \text{ matrix}\right]$$

$$\boldsymbol{\psi} = \left(\boldsymbol{\psi}_1, \boldsymbol{\psi}_2, ..., \boldsymbol{\psi}_I\right)\left[(I\times G\times 4) \text{ matrix}\right]$$

With this, we can perform the normalisation as follows:

$$\begin{aligned}\boldsymbol{x}_N = {} & I\left(\boldsymbol{x} \,.\!< \boldsymbol{\psi}[.,.,1]\right)*0 + I\left(\boldsymbol{\psi}[.,.,1] \,.\!< \boldsymbol{x} \,.\!< \boldsymbol{\psi}[.,.,2]\right)*\left(\frac{\boldsymbol{x}-\boldsymbol{\psi}[.,.,1]}{\boldsymbol{\psi}[.,.,2]-\boldsymbol{\psi}[.,.,1]}\right) \\ & + I\left(\boldsymbol{\psi}[.,.,2] \,.\!< \boldsymbol{x} \,.\!< \boldsymbol{\psi}[.,.,3]\right)*1 + I\left(\boldsymbol{\psi}[.,.,3] \,.\!< \boldsymbol{x} \,.\!< \boldsymbol{\psi}[.,.,4]\right)*\left(\frac{\boldsymbol{\psi}[.,.,4]-\boldsymbol{x}}{\boldsymbol{\psi}[.,.,4]-\boldsymbol{\psi}[.,.,3]}\right) \\ & + I\left(\boldsymbol{\psi}[.,.,4] \,.\!< \boldsymbol{x}\right)*0\end{aligned}$$

where $\boldsymbol{x}_N$ is the normalised attribute matrix of size $(I\times G)$, $I(.)$ is the indicator function which returns a value of 1 if the condition is true otherwise 0, and $.<$ is an element-by-element comparison operator.

Next, using $\boldsymbol{x}_N$ and $\boldsymbol{\mu}$, we evaluate the $CI$ for each alternative using equation A.2 and write equation A.1 in the matrix for as follows:

$$\boldsymbol{U} = \boldsymbol{CI} + \boldsymbol{\varepsilon} \qquad \text{(A.3)}$$

Where $\boldsymbol{CI} = \left(CI_1, CI_2, ..., CI_I\right)\left[(I\times 1) \text{ vector}\right]$. Thus, we can write the distribution of $\boldsymbol{U}$ as $\boldsymbol{U} \sim MVN_{(I\times I)}\left[\boldsymbol{CI}, \boldsymbol{\Lambda}\right]$ where $\boldsymbol{\Lambda}$ is the covariance matrix of $\boldsymbol{\varepsilon}$.

### A.1.2 Estimation

Since, only the difference in utility matters, we work with utility differences. We specifically subtract the utility of the chosen alternative from utilities of all non-chosen alternatives. Moreover, top left element of the differenced error covariance matrix ( $\tilde{\mathbf{\Lambda}}$ ) is fixed to 1 to set the utility scale for identifiability (Train, 2009). Thus, for $I$ alternative, only $\left[I*(I-1)*0.5\right]-1$ covariance elements are identifiable. Further, since all the differenced error covariance matrices must originate from the same undifferenced error covariance matrix ( $\mathbf{\Lambda}$ ), we specify matrix $\mathbf{\Lambda}$ as follows: $\mathbf{\Lambda}=\begin{bmatrix}0 & 0\\ 0 & \tilde{\mathbf{\Lambda}}\end{bmatrix}$. To perform utility difference, we construct a matrix $\mathbf{M}$ of size $[(I-I)\times I]$ using the following pseudo-code:

```
Iden_mat = 1_{I-1}
O_neg    = -1*ones(I-1,1)
if(i_m == 1)
   M = O_neg~Iden_mat
elseif(i_m == I)
   M = Iden_mat ~ O_neg
else
   M = Iden_mat[.,1:i_m-1] ~ O_neg~Iden_mat[.,i_m:I-1]
end
```

where " ~ " refers to horizontal concatenation and $i_m$ is the chosen alternative.

Using $\mathbf{M}$, we can write the distribution of utility differences $\bar{\boldsymbol{U}} \sim MVN_{(I-I)}\left(\tilde{\mathbf{B}},\tilde{\mathbf{\Theta}}\right)$ , where $\tilde{\mathbf{B}}=\mathbf{M}*\boldsymbol{CI}$, and $\tilde{\mathbf{\Theta}}=\mathbf{M}*\mathbf{\Lambda}*\mathbf{M}'$. Thus, the likelihood of the decision-maker $n$ can be written as: $L_n(\boldsymbol{\theta})=\int_{-\infty}^{\tilde{\mathbf{B}}} f_{(I-I)}\left(\boldsymbol{r}\,\middle|\,\tilde{\mathbf{B}}\ ,\tilde{\mathbf{\Theta}}\right)d\mathbf{r}$ . Thus, the constrained likelihood maximization problem becomes:

$$\max_{\boldsymbol{\theta}} \sum_{n=1}^{N} Log\left(L_n(\boldsymbol{\theta})\right) \tag{A.4}$$

***Such that for each alternative*** $\forall i$

$$\sum_{H \subseteq A_G} m(H) = 1; \quad \text{where } A_G = \{x_1, x_2, ..., x_G\}$$

$$\sum_{H \subseteq A_G \setminus g} m(H \cup k) \geq 0 \qquad \forall g \subseteq A_G, \forall k \subseteq A_G; \forall i$$

where $A_G \setminus g$ represents collection of all attributes except the $g^{th}$ attribute (A.5)

$\cup$ represents the union of two sets

$m(.)$ is the Möbius representation of $\mu(.)$

$$m(H) = \sum_{F \subseteq H} (-1)^{|H \setminus F|} \mu(F)$$

Further, the one-to-one mapping between $\mu(.)$ and $m(.)$ is as follows:

$$\mu(F) = \sum_{H \subseteq F} m(H)$$

We convert fuzzy-measures (matrix $\boldsymbol{\mu}$) into their corresponding Möbius parameters (matrix $\boldsymbol{m}$) and solve the above constrained optimization problem. The decision variables in the constrained maximisation problem are $\boldsymbol{\theta} = \left[\text{Vech}(\boldsymbol{m}), \text{Vech}(\boldsymbol{\psi}), \text{Vech}(\tilde{\boldsymbol{\Lambda}})\right]$, where Vech(.) operator vectorises the unique element of a matrix. Readers will note that the number of constraints does not depend on the number of alternatives in generic CI-based indirect utility specification, but they grow linearly with the number of alternatives in alternative-specific CI.

The likelihood function involves computation of a $(I-1)$ dimensional multivariate normal cumulative density function (MVNCDF) for each decision-maker. One can use Geweke, Hajivassiliou and Keane (GHK) simulator (Geweke, 1991; Hajivassiliou et al., 1992; Keane, 1994; Genz, 1992) or analytical approximation methods (Bhat, 2011; Bhat, 2018) to accurately evaluate the multivariate normal cumulative distribution function (MVNCDF). For approximate computation of the MVNCDF function, we use GHK simulator with Halton Draws (Bhat, 2014; Train, 2009).

**Positive definiteness of error-differenced covariance matrix:**

In order to maintain the positive definiteness of the error covariance matrix, we work with the Cholesky decomposition. Since the first element of error differenced covariance matrix is fixed to 1, we use the following parametrisation on the Cholesky decomposition.

Let $\mathbf{LL}' = \tilde{\mathbf{\Lambda}}$, where $\mathbf{L}$ is the lower traingular Cholesky matrix. To derive $\mathbf{L}_p$ from $\mathbf{L}$, we first compute $a_i = \left[1 + \left(\mathbf{L}[i,1:i-1]\right)^2\right]^{0.5}$ $\forall i \geq 2$. Then, we parametrize all the non-diagonal elements of the $i^{th}$ row as $\mathbf{L}_p[i,r] = \dfrac{\mathbf{L}[i,r]}{a_i}$ $\forall r = 1$ to $i-1$ and the diagonal element as $\mathbf{L}_p[i,i] = \dfrac{1}{a_i}$.

For example: consider a differenced error-covariance matrix of three alternatives as follow:

$\tilde{\mathbf{\Lambda}} = \begin{bmatrix} 1.0 & 0.5 \\ 0.5 & 1.2 \end{bmatrix}$. Then, the corresponding lower triangular Cholesky matrix can be written as follow: $\mathbf{L} = \begin{bmatrix} 1.0 & 0 \\ 0.5 & 0.98 \end{bmatrix}$. Then, we obtain $a_2 = \left[1 + (0.5)^2\right]^{0.5} = 1.12$. Therefore, $\mathbf{L}_p$ can be parameterised as follow: $\mathbf{L}_p = \begin{bmatrix} 1.0 & 0 \\ 0.50/1.12 & 1.00/1.12 \end{bmatrix} = \begin{bmatrix} 1.0 & 0 \\ 0.45 & 0.89 \end{bmatrix}$

## A.2 Data Generating Process

### A.2.1 Four-attribute configuration

Alternative-specific intercepts: $\begin{bmatrix} ASC_1 \\ ASC_2 \\ ASC_3 \\ ASC_4 \\ ASC_5 \end{bmatrix} = \begin{bmatrix} 0 \\ -0.7 \\ -0.6 \\ -0.5 \\ -0.4 \end{bmatrix}$

Choquet integral parameters for MNP-CI:

$\mu(1) = 0.3, \mu(2) = 0.25, \mu(3) = 0.2, \mu(4) = 0.1,$

$\mu(12) = 0.58, \mu(13) = 0.53, \mu(14) = 0.44, \mu(23) = 0.49, \mu(24) = 0.36, \mu(34) = 0.33,$

$\mu(123) = 0.79, \mu(124) = 0.68, \mu(134) = 0.64, \mu(234) = 0.59, \mu(1234) = 1.0$

Mean effect parameters for MNP-WS: $\beta(1) = 0.3, \beta(2) = 0.25, \beta(3) = 0.2, \beta(4) = 0.1$

Cut-off parameters:

| **Explanatory variables** | **Cut-off type** | **Cut-off points** | | | |
|---|---|---|---|---|---|
| | | *a* | *b* | *c* | *d* |
| 1 | Half-triangular | 3.0 | 7.0 | | |
| 2 | Half-triangular | 3.5 | 6.5 | | |
| 3 | Trapezoidal | 2.0 | 4.0 | 6.0 | 7.0 |
| 4 | Trapezoidal | 3.5 | 5.5 | 7.5 | 8.5 |

Shapley values and Interaction indices:

| **Explanatory variable** | **Shapley value** | **Explanatory variable pair** | **Interaction index** |
|---|---|---|---|
| $S(1)$ | 0.338 | $I(12)$ | 0.035 |
| $S(2)$ | 0.285 | $I(13)$ | 0.030 |
| $S(3)$ | 0.242 | $I(14)$ | 0.050 |
| $S(4)$ | 0.135 | $I(23)$ | 0.045 |
| | | $I(24)$ | 0.025 |
| | | $I(34)$ | 0.040 |

Error structure: $\begin{bmatrix} 1.0 & & & \\ 0.5 & 1.1 & & \\ 0.5 & 0.5 & 1.2 & \\ 0.5 & 0.5 & 0.5 & 1.3 \end{bmatrix}$

### A.2.2 Six-attribute configuration

Alternative-specific intercepts: $\begin{bmatrix} ASC_1 \\ ASC_2 \\ ASC_3 \\ ASC_4 \\ ASC_5 \end{bmatrix} = \begin{bmatrix} 0 \\ -0.7 \\ -0.6 \\ -0.5 \\ -0.4 \end{bmatrix}$

Choquet integral parameters for MNP-CI:

$\mu(1)=0.17, \mu(2)=0.18, \mu(3)=0.20, \mu(4)=0.16, \mu(5)=0.19, \mu(6)=0.18,$
$\mu(12)=0.33, \mu(13)=0.35, \mu(14)=0.31, \mu(15)=0.34, \mu(16)=0.33,$
$\mu(23)=0.36, \mu(24)=0.32, \mu(25)=0.35, \mu(26)=0.34,$
$\mu(34)=0.34, \mu(35)=0.37, \mu(36)=0.36,$
$\mu(45)=0.33, \mu(46)=0.32, \mu(56)=0.35,$
$\mu(123)=0.51, \mu(124)=0.47, \mu(125)=0.50, \mu(126)=0.49, \mu(134)=0.49, \mu(135)=0.52,$
$\mu(136)=0.51, \mu(145)=0.48, \mu(146)=0.47, \mu(156)=0.50,$
$\mu(234)=0.50, \mu(235)=0.53, \mu(236)=0.52, \mu(245)=0.49, \mu(246)=0.48, \mu(256)=0.51,$
$\mu(345)=0.51, \mu(346)=0.50, \mu(356)=0.53, \mu(456)=0.49,$
$\mu(1234)=0.65, \mu(1235)=0.68, \mu(1236)=0.67, \mu(1245)=0.64, \mu(1246)=0.63, \mu(1256)=0.66,$
$\mu(1345)=0.66, \mu(1346)=0.65, \mu(1356)=0.68, \mu(1456)=0.64,$
$\mu(2345)=0.67, \mu(2346)=0.66, \mu(2356)=0.69, \mu(2456)=0.65, \mu(3456)=0.67,$
$\mu(12345)=0.82, \mu(12346)=0.81, \mu(12356)=0.84, \mu(12456)=0.80, \mu(13456)=0.82,$
$\mu(23456)=0.83, \mu(123456)=1.00$

Mean effect parameters for MNP-WS:

$\beta(1)=0.17, \beta(2)=0.18, \beta(3)=0.20, \beta(4)=0.16, \beta(5)=0.19, \beta(6)=0.18$

Cut-off parameters:

| Explanatory variables | Cut-off type | Cut-off points | | | |
|---|---|---|---|---|---|
| | | *a* | *b* | *c* | *d* |
| 1 | Half-triangular | 3.0 | 7.0 | | |
| 2 | Half-triangular | 3.5 | 6.5 | | |
| 3 | Trapezoidal | 2.0 | 4.0 | 6.0 | 7.0 |
| 4 | Trapezoidal | 3.5 | 5.5 | 7.5 | 8.5 |
| 5 | Half-triangular | 3.3 | 6.8 | | |
| 6 | Trapezoidal | 2.5 | 5.0 | 6.5 | 7.5 |

Shapley values and Interaction indices:

| Explanatory variable | Shapley value | Explanatory variable pair | Interaction index |
|---|---|---|---|
| $S(1)$ | 0.157 | $I(12)$ | 0.00 |
| $S(2)$ | 0.167 | $I(13)$ | 0.00 |
| $S(3)$ | 0.187 | $I(14)$ | 0.00 |
| $S(4)$ | 0.147 | $I(15)$ | 0.00 |
| $S(5)$ | 0.177 | $I(16)$ | 0.00 |
| $S(6)$ | 0.167 | $I(23)$ | 0.00 |
| | | $I(24)$ | 0.00 |
| | | $I(25)$ | 0.00 |
| | | $I(26)$ | 0.00 |
| | | $I(34)$ | 0.00 |
| | | $I(35)$ | 0.00 |
| | | $I(36)$ | 0.00 |
| | | $I(45)$ | 0.00 |
| | | $I(46)$ | 0.00 |
| | | $I(56)$ | 0.00 |

Error structure: $\begin{bmatrix} 1.0 & & & \\ 0.5 & 1.1 & & \\ 0.5 & 0.5 & 1.2 & \\ 0.5 & 0.5 & 0.5 & 1.3 \end{bmatrix}$

# Online Appendix

## A Multinomial Probit Model with Choquet Integral and Attribute Cut-offs

## S.1 Example of CI computation

As an illustration, we show how to calculate CI for a case of three attributes.

Let $\boldsymbol{X} = \{x_1, x_2, x_3\}$.I.e, $G = 3$

Further, discrete fuzzy measures have the following configuration:

$$\mu(\phi) = 0;\ \mu(x_1) = 0.2;\ \mu(x_2) = 0.3;\ \mu(x_3) = 0.1;$$
$$\mu(x_1 x_2) = 0.687;\ \mu(x_1 x_3) = 0.362;\ \mu(x_2 x_3) = 0.493;$$
$$\mu(x_1 x_2 x_3) = 1$$

Next, the observed value of three attributes is as follow: $\{x_1 = 0.3, x_2 = 0.1, x_3 = 1\}$.

We observe that $x_3 > x_1 > x_2$

Therefore $h(x_{\pi_1}) = x_3$ $h(x_{\pi_2}) = x_1$ $h(x_{\pi_3}) = x_2$ and

$A_0 = \phi$ $A_1 = \{x_3\}$ $A_2 = \{x_3, x_1\}$ $A_3 = \{x_3, x_1, x_2\}$. Thus,

$$\begin{aligned}
CI &= h(x_{\pi_1})(\mu(A_1) - \mu(A_0)) + h(x_{\pi_2})(\mu(A_2) - \mu(A_1)) + h(x_{\pi_3})(\mu(A_3) - \mu(A_2)) \\
&= x_3(\mu(x_3) - \mu(\phi)) + x_1(\mu(x_3 x_1) - \mu(x_3)) + x_2(\mu(x_3 x_1 x_2) - \mu(x_3 x_1)) \\
&= 1(0.1 - 0) + 0.3(0.362 - 0.1) + 0.1(1 - 0.362) \\
&= 0.242
\end{aligned}$$

# S.2 Functions Nested by Choquet Integral

## S.2.1 Choquet Integral as Weighted Sum:

### *Calculation of Choquet Integral when fuzzy measures are additive*

Let $\boldsymbol{X} = \{x_1, x_2, x_3\}$ and $\mu(ab) = \mu(a) + \mu(b)$ (i.e, they are additive)

$\mu(\phi) = 0;\ \mu(x_1) = 0.4;\ \mu(x_2) = 0.45;\ \mu(x_3) = 0.15;\ \mu(x_1x_2) = 0.85;\ \mu(x_1x_3) = 0.55;\ \mu(x_2x_3) = 0.60$

Observed Value $\{x_1 = 0.3, x_2 = 0.1, x_3 = 1\}$

Here $x_3 > x_1 > x_3$

So, $h(x_{\pi_1}) = x_3; h(x_{\pi_2}) = x_1; h(x_{\pi_3}) = x_2$

$A_0 = \phi\ \ A_1 = \{x_3\}\ \ A_2 = \{x_3, x_1\}\ \ A_3 = \{x_3, x_1, x_2\}$

$$CI = \pi_1(\mu(A_1) - \mu(A_0)) + \pi_2(\mu(A_2) - \mu(A_1)) + \pi_3(\mu(A_3) - \mu(A_2))$$
$$= x_3(\mu(x_3) - \mu(\phi)) + x_1(\mu(x_3, x_1) - \mu(x_3)) + x_2(\mu(x_3, x_1, x_2) - \mu(x_3, x_1))$$
$$= 1(0.15 - 0) + 0.3(0.55 - 0.15) + 0.1(1 - 0.55)$$
$$= 0.315$$

### *Calculation of weighted sum using additive fuzzy measure weights*

Let $\boldsymbol{X} = \{x_1, x_2, x_3\}$ and $\mu(ab) = \mu(a) + \mu(b)$ (i.e, they are additive)

$\mu(\phi) = 0;\ \mu(x_1) = 0.4;\ \mu(x_2) = 0.45;\ \mu(x_3) = 0.15;\ \mu(x_1x_2) = 0.85;\ \mu(x_1x_3) = 0.55;\ \mu(x_2x_3) = 0.60$

Observed Value $\{x_1 = 0.3, x_2 = 0.1, x_3 = 1\}$

$$WS = x_1\mu(x_1) + x_2\mu(x_2) + x_3\mu(x_3)$$
$$= 0.3 * 0.4 + 0.1 * 0.45 + 1 * 0.15$$
$$= 0.315$$

**S.2.2 Choquet Integral as Ordered Weighted Sum (OWS):**

***Calculation of Choquet Integral when fuzzy measures are symmetric***

Let $\boldsymbol{X} = \{x_1, x_2, x_3\}$ and $\mu(A) = f(|A|)$ (i.e, they are symmetric (function of cardinality of set))

Let's assume $\mu(A) = 0.333*(|A|)$

$\mu(\phi) = 0;\ \mu(x_1), \mu(x_2), \mu(x_3) = 0.333;\ ;\ \mu(x_1x_2), \mu(x_1x_3), \mu(x_2x_3) = 0.666;\ \mu(x_1x_2x_3) = 0.999$

Observed Value $\{x_1 = 0.3, x_2 = 0.1, x_3 = 1\}$

Here $x_3 > x_1 > x_3$

So, $h(x_{\pi_1}) = x_3; h(x_{\pi_2}) = x_1; h(x_{\pi_3}) = x_2$

$A_0 = \phi\ \ A_1 = \{x_3\}\ \ A_2 = \{x_3, x_1\}\ \ A_3 = \{x_3, x_1, x_2\}$

$$
\begin{aligned}
CI &= \pi_1(\mu(A_1) - \mu(A_0)) + \pi_2(\mu(A_2) - \mu(A_1)) + \pi_3(\mu(A_3) - \mu(A_2)) \\
&= x_3(\mu(x_3) - \mu(\phi)) + x_1(\mu(x_3, x_1) - \mu(x_3)) + x_2(\mu(x_3, x_1, x_2) - \mu(x_3, x_1)) \\
&= 1(0.333 - 0) + 0.3(0.666 - 0.333) + 0.1(0.999 - 0.666) \\
&= 0.4662
\end{aligned}
$$

***Calculation of ordered weighted sum using symmetric fuzzy measure weights***

Let $\boldsymbol{X} = \{x_1, x_2, x_3\}$ and $\mu(A) = f(|A|)$ (i.e, they are symmetric (function of cardinality of set))

Let's assume $\mu(A) = 0.333*(|A|)$

$\mu(\phi) = 0;\ \mu(x_1), \mu(x_2), \mu(x_3) = 0.333;\ ;\ \mu(x_1x_2), \mu(x_1x_3), \mu(x_2x_3) = 0.666;\ \mu(x_1x_2x_3) = 0.999$

Observed Value $\{x_1 = 0.3, x_2 = 0.1, x_3 = 1\}$

Here $x_3 > x_1 > x_3$

$$
\begin{aligned}
OWS &= x_3\mu(x_3) + x_1\mu(x_1) + x_2\mu(x_2) \\
&= 1(0.333) + 0.3(0.333) + 0.1(0.333) \\
&= 0.4662
\end{aligned}
$$

### S.2.3 Choquet Integral as Minimum or Maximum of Attributes

Let $\boldsymbol{X} = \{x_1, x_2, x_3\}$

$\mu(\phi) = 0;\ \mu(x_1) = 0;\ \mu(x_2) = 0;\ \mu(x_3) = 1;\ \mu(x_1x_2) = 0;\ \mu(x_1x_3) = 1;\ \mu(x_2x_3) = 1; \mu(x_1x_2x_3) = 1$

The above configuration of fuzzy-measures are additive

Observed Value $\{x_1 = 0.3, x_2 = 0.1, x_3 = 1\}$

Here $x_3 > x_1 > x_3$

So, $h(x_{\pi_1}) = x_3; h(x_{\pi_2}) = x_1; h(x_{\pi_3}) = x_2$

$A_0 = \phi\ \ A_1 = \{x_3\}\ \ A_2 = \{x_3, x_1\}\ \ A_3 = \{x_3, x_1, x_2\}$

$$\begin{aligned} CI &= \pi_1(\mu(A_1) - \mu(A_0)) + \pi_2(\mu(A_2) - \mu(A_1)) + \pi_3(\mu(A_3) - \mu(A_2)) \\ &= x_3(\mu(x_3) - \mu(\phi)) + x_1(\mu(x_3, x_1) - \mu(x_3)) + x_2(\mu(x_3, x_1, x_2) - \mu(x_3, x_1)) \\ &= 1(1-0) + 0.3(1-1) + 0.1(1-1) \\ &= 1\ (\text{Maximum of attributes}) \end{aligned}$$

Similarly,for the following configuration, we can obtain minimum of attributes

$\mu(\phi) = 0;\ \mu(x_1) = 0;\ \mu(x_2) = 1;\ \mu(x_3) = 0;\ \mu(x_1x_2) = 1;\ \mu(x_1x_3) = 0;\ \mu(x_2x_3) = 1; \mu(x_1x_2x_3) = 1$

$$\begin{aligned} CI &= \pi_1(\mu(A_1) - \mu(A_0)) + \pi_2(\mu(A_2) - \mu(A_1)) + \pi_3(\mu(A_3) - \mu(A_2)) \\ &= x_3(\mu(x_3) - \mu(\phi)) + x_1(\mu(x_3, x_1) - \mu(x_3)) + x_2(\mu(x_3, x_1, x_2) - \mu(x_3, x_1)) \\ &= 1(0-0) + 0.3(0-0) + 0.1(1-0) \\ &= 0.1\ (\text{Minimum of attributes}) \end{aligned}$$

## S.3 Example of Mapping between Möbius Transform and Fuzzy Measures

Consider there are 4 attributes $g = \{1, 2, 3, 4\}$ and $\mu(.)$ and $m(.)$ represent the fuzzy measure and Möbius parameters, respectively. Then the equality and inequality constraints can be written using Möbius parameters (and their implied fuzzy measures conditions) as follow:

**Equality constraint**

$$m(1) + m(2) + m(3) + m(4) + m(12) + m(13) + m(14) + m(23) + m(24) + m(34) + m(123) + m(124) + m(134) + m(234) + m(1234) = 1$$

This constraint implies that $\mu(1234) = 1$

**Inequality constraints**

| | |
|---|---|
| $m(1) \geq 0$ | $\mu(1) \geq 0$ |
| $m(2) \geq 0$ | $\mu(2) \geq 0$ |
| $m(3) \geq 0$ | $\mu(3) \geq 0$ |
| $m(4) \geq 0$ | $\mu(4) \geq 0$ |
| $m(1) + m(12) \geq 0$ | $\mu(12) - \mu(2) \geq 0$ |
| $m(1) + m(13) \geq 0$ | $\mu(13) - \mu(3) \geq 0$ |
| $m(1) + m(14) \geq 0$ | $\mu(14) - \mu(4) \geq 0$ |
| $m(2) + m(12) \geq 0$ | $\mu(12) - \mu(1) \geq 0$ |
| $m(2) + m(23) \geq 0$ | $\mu(23) - \mu(3) \geq 0$ |
| $m(2) + m(24) \geq 0$ | $\mu(24) - \mu(4) \geq 0$ |
| $m(3) + m(13) \geq 0$ | $\mu(13) - \mu(1) \geq 0$ |
| $m(3) + m(23) \geq 0$ | $\mu(23) - \mu(2) \geq 0$ |
| $m(3) + m(34) \geq 0$ | $\mu(34) - \mu(4) \geq 0$ |
| $m(4) + m(14) \geq 0$ | $\mu(14) - \mu(1) \geq 0$ |
| $m(4) + m(24) \geq 0$ | $\mu(24) - \mu(2) \geq 0$ |
| $m(4) + m(34) \geq 0$ | $\mu(34) - \mu(3) \geq 0$ |
| $m(1) + m(12) + m(13) + m(123) \geq 0$ | $\mu(123) - \mu(23) \geq 0$ |
| $m(1) + m(12) + m(14) + m(124) \geq 0$ | $\mu(124) - \mu(24) \geq 0$ |
| $m(1) + m(13) + m(14) + m(134) \geq 0$ | $\mu(134) - \mu(34) \geq 0$ |
| $m(2) + m(12) + m(23) + m(123) \geq 0$ | $\mu(123) - \mu(13) \geq 0$ |
| $m(2) + m(12) + m(24) + m(124) \geq 0$ | $\mu(124) - \mu(14) \geq 0$ |
| $m(2) + m(23) + m(24) + m(234) \geq 0$ | $\mu(234) - \mu(34) \geq 0$ |
| $m(3) + m(13) + m(23) + m(123) \geq 0$ | $\mu(123) - \mu(12) \geq 0$ |
| $m(3) + m(13) + m(34) + m(134) \geq 0$ | $\mu(134) - \mu(14) \geq 0$ |
| $m(3) + m(23) + m(34) + m(234) \geq 0$ | $\mu(234) - \mu(24) \geq 0$ |
| $m(4) + m(14) + m(24) + m(124) \geq 0$ | $\mu(124) - \mu(12) \geq 0$ |
| $m(4) + m(14) + m(34) + m(134) \geq 0$ | $\mu(134) - \mu(13) \geq 0$ |
| $m(4) + m(24) + m(34) + m(234) \geq 0$ | $\mu(234) - \mu(23) \geq 0$ |
| $m(1) + m(12) + m(13) + m(123) + m(14) + m(124) + m(134) + m(1234) \geq 0$ | $\mu(1234) - \mu(234) \geq 0$ |
| $m(2) + m(12) + m(23) + m(123) + m(24) + m(124) + m(234) + m(1234) \geq 0$ | $\mu(1234) - \mu(134) \geq 0$ |
| $m(3) + m(13) + m(23) + m(123) + m(34) + m(134) + m(234) + m(1234) \geq 0$ | $\mu(1234) - \mu(124) \geq 0$ |
| $m(4) + m(14) + m(24) + m(124) + m(34) + m(134) + m(234) + m(1234) \geq 0$ | $\mu(1234) - \mu(123) \geq 0$ |

## S.4 Fuzzy Membership Functions for Attribute Cut-offs

Attributes with positive marginal utility such as the number of seats, doors, storage area in a vehicle choice scenario can be represented using the following half triangular function (see Figure O.1):

$$x_N = \begin{cases} 0 & x \le a \\ \dfrac{x-a}{b-a} & a < x \le b \\ 1 & x > b \end{cases} \tag{O.1}$$

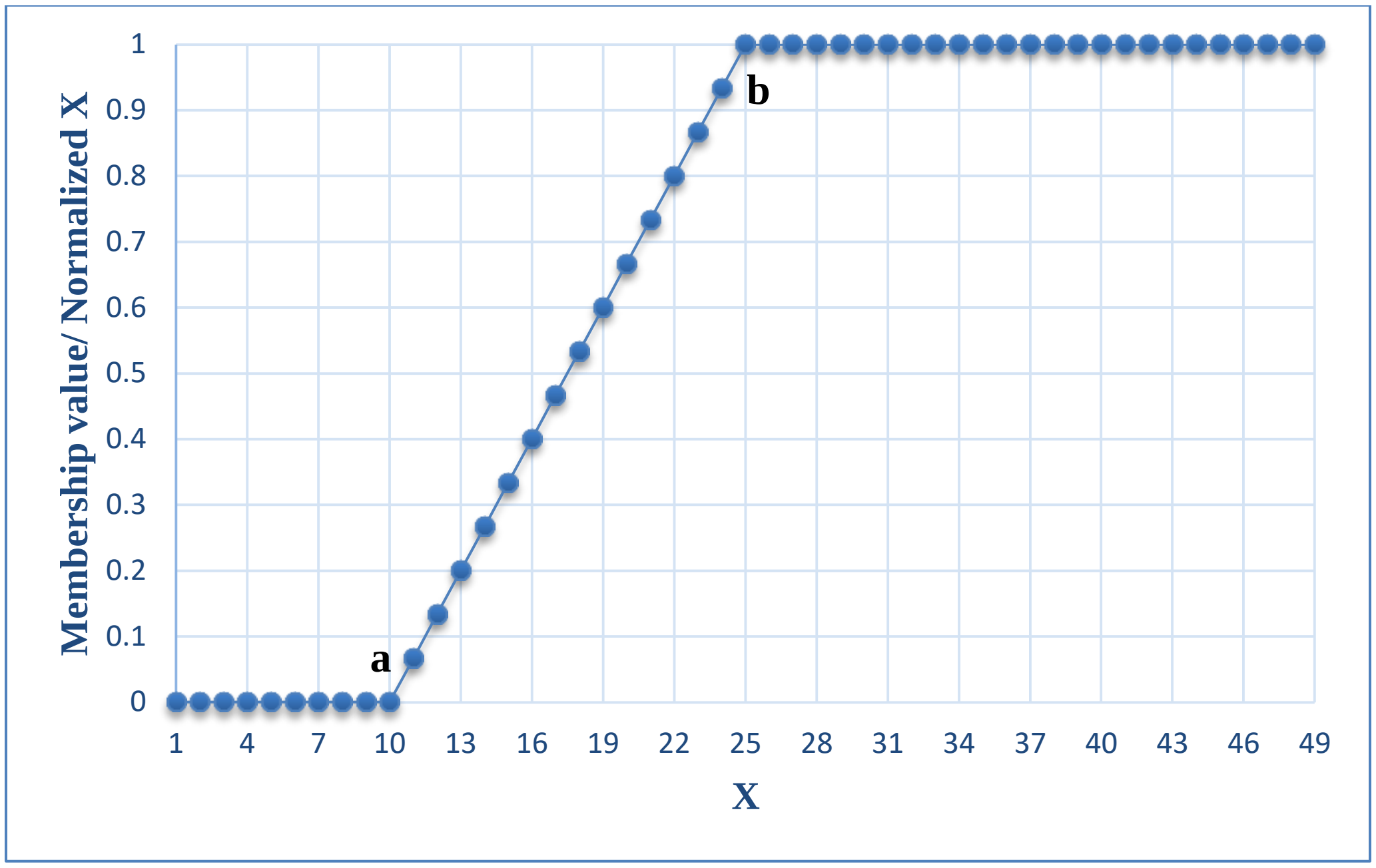

**Figure O.1:** Two-point cut-off graph for attributes with negative marginal utility (i.e., half-triangular function)

Similarly, variables such as driving range in an electric vehicle choice can be represented by the following trapezoidal membership function (see Figure O.2):

$$x_N = \begin{cases} 0 & x \le a; x > d \\ \dfrac{x-a}{b-a} & a < x \le b \\ 1 & b < x \le c \\ \dfrac{d-x}{d-c} & c < x \le d \end{cases} \tag{O.2}$$

It is worth noting that the triangular membership function can easily be represented by using a trapezoidal function after imposing equality constraint on $b$ and $c$.

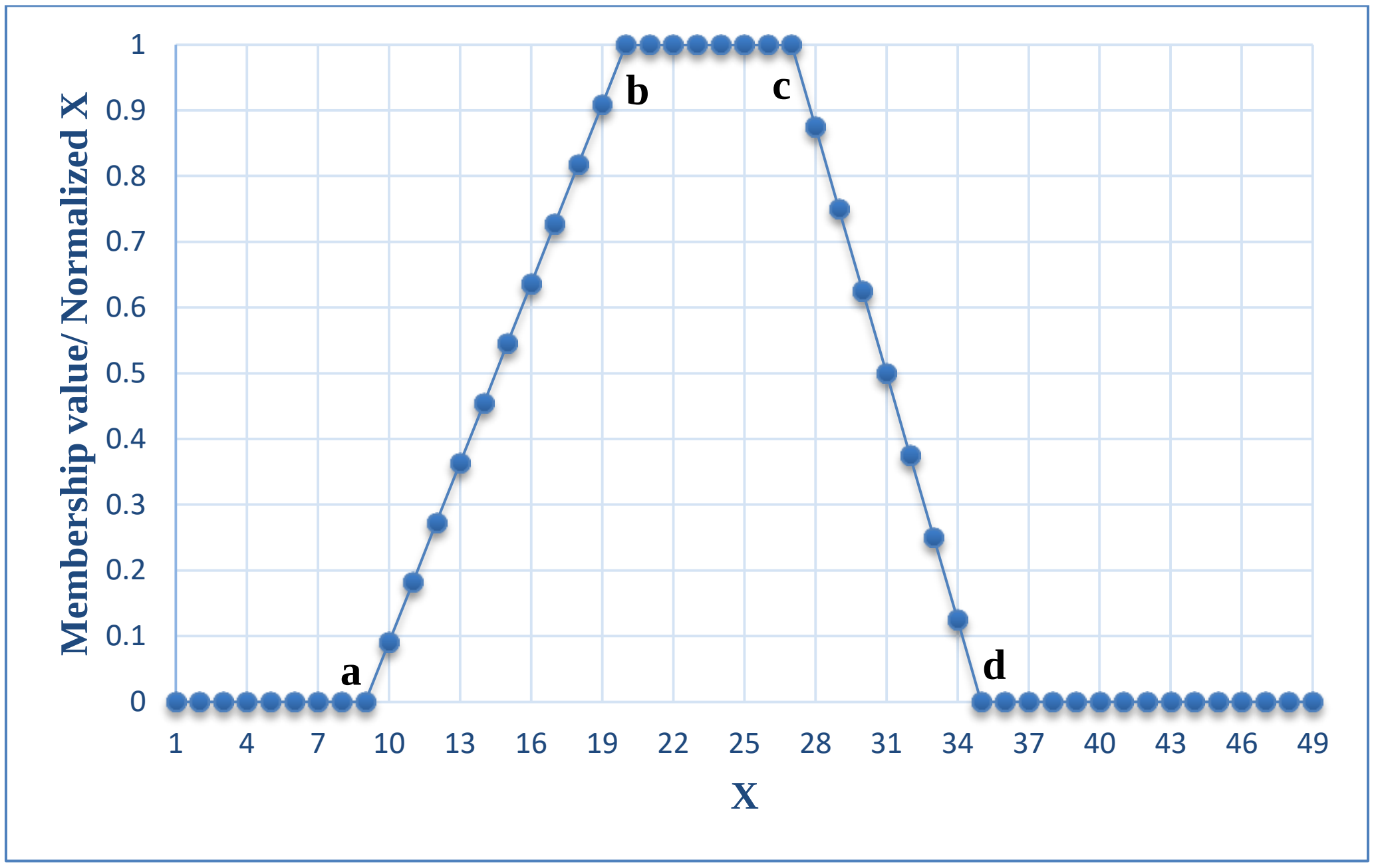

**Figure O.2:** Four-point cut-off graph (i.e., trapezoidal membership function)

## S.5 Operationalisation and Interpretation of Attribute Cut-offs in CI

Consider that an individual needs to make a binary choice (choose or not choose) for a Mobility-on-Demand (MoD) service in three scenarios, where a scenario is characterised with three attributes in-vehicle travel time per km (IVTT/Km), out-of-vehicle travel time per km (OVTT/Km), and cost per km (Cost/Km)) as shown below in Table O.1.

**Table O.1:** MoD choice scenario

| Scenario | IVTT/Km | OVTT/Km | Cost/Km |
|---|---|---|---|
| 1 | 5 | 3 | 1.6 |
| 2 | 4 | 4 | 1.6 |
| 3 | 4 | 2 | 2 |

Further, we assume that the individual considers a half-triangular fuzzy membership function for all three attributes with cut-off points $a$ and $b$ as follow:

$$\text{IVTT/Km}[a,b]=(2.5,4.5),\text{OVTT/Km}[a,b]=(1.5,3.5),\text{Cost/Km}[a,b]=(1.0,1.9)$$

Using the attribute cut-offs, we obtain the normalized value for three attributes in all scenarios, as shown in Table O.2.

**Table O.2:** MoD choice scenario normalized value

| Scenario | IVTT/Km | OVTT/Km | Cost/Km |
|---|---|---|---|
| 1 | 0 | 0.25 | 0.33 |
| 2 | 0.25 | 0 | 0.33 |
| 3 | 0.25 | 0.75 | 0 |

Further, we assume the following configuration for fuzzy measures:

$$\mu(\phi)=0;\ \mu(\text{IVTT/Km})=0.087;\ \mu(\text{OVTT/Km})=0.21;\ \mu(\text{Cost/Km})=0.443;$$
$$\mu(\text{IVTT/Km, OVTT/Km})=0.382;\ \mu(\text{IVTT/Km, Cost/Km})=0.595;\ \mu(\text{OVTT/Km,Cost/Km})=0.653;$$
$$\mu(\text{IVTT/Km, OVTT/Km,Cost/Km})=1.00$$

Then, we can calculate the CI value of three scenarios using the normalised value of attributes (Table O.2) and the fuzzy measures as follows:

$$CI(1) = \text{Cost/Km}\left(\mu(\text{Cost/Km}) - \mu(\phi)\right) + \text{OVTT/Km}\left(\mu(\text{OVTT/Km, Cost/Km}) - \mu(\text{Cost/Km})\right)$$
$$+ \text{IVTT/Km}\left(\mu(\text{IVTT/Km, OVTT/Km, Cost/Km}) - \mu(\text{OVTT/Km, Cost/Km})\right)$$
$$= 0.33(0.443 - 0) + 0.25(0.653 - 0.443) + 0(1 - 0.653) \; = 0.1987$$
$$CI(2) = \text{Cost/Km}\left(\mu(\text{Cost/Km}) - \mu(\phi)\right) + \text{IVTT/Km}\left(\mu(\text{IVTT/Km, Cost/Km}) - \mu(\text{Cost/Km})\right)$$
$$+ \text{OVTT/Km}\left(\mu(\text{IVTT/Km, OVTT/Km, Cost/Km}) - \mu(\text{IVTT/Km, Cost/Km})\right)$$
$$= 0.33(0.443 - 0) + 0.25(0.595 - 0.443) + 0(1 - 0.595) \; = 0.1842$$
$$CI(3) = \text{OVTT/Km}\left(\mu(\text{OVTT/Km}) - \mu(\phi)\right) + \text{IVTT/Km}\left(\mu(\text{IVTT/Km, OVTT/Km}) - \mu(\text{OVTT/Km})\right)$$
$$+ \text{Cost/Km}\left(\mu(\text{IVTT/Km, OVTT/Km, Cost/Km}) - \mu(\text{IVTT/Km, OVTT/Km})\right)$$
$$= 0.75(0.21 - 0) + 0.25(0.382 - 0.21) + 0(1 - 0.382) \; = 0.2005$$

Finally, assuming a probit choice probability kernel, we obtain the probability of choosing MoD as shown in Table O.3.

**Table O.3:** Utility and corresponding probability of choosing MoD

| **Scenario** | **Utility** | **Probability of choice** |
|---|---|---|
| 1 | 0.1987 | 0.336 |
| 2 | 0.1842 | 0.327 |
| 3 | 0.2005 | 0.337 |

It is worth noting that each scenario sets one of the attribute values to be zero after normalization (see Table O.2) because the realized value of attribute goes beyond the upper limit. This can be viewed as a situation where the contribution of the attribute to the utility of an alternative beyond a certain attribute threshold does not change. For instance, zero value for IVTT/km in first choice scenario indicates that the value of 5 for IVTT/km cause the same disutility as the value of 4.5.

## S.6 Additional Results of Monte Carlo Study

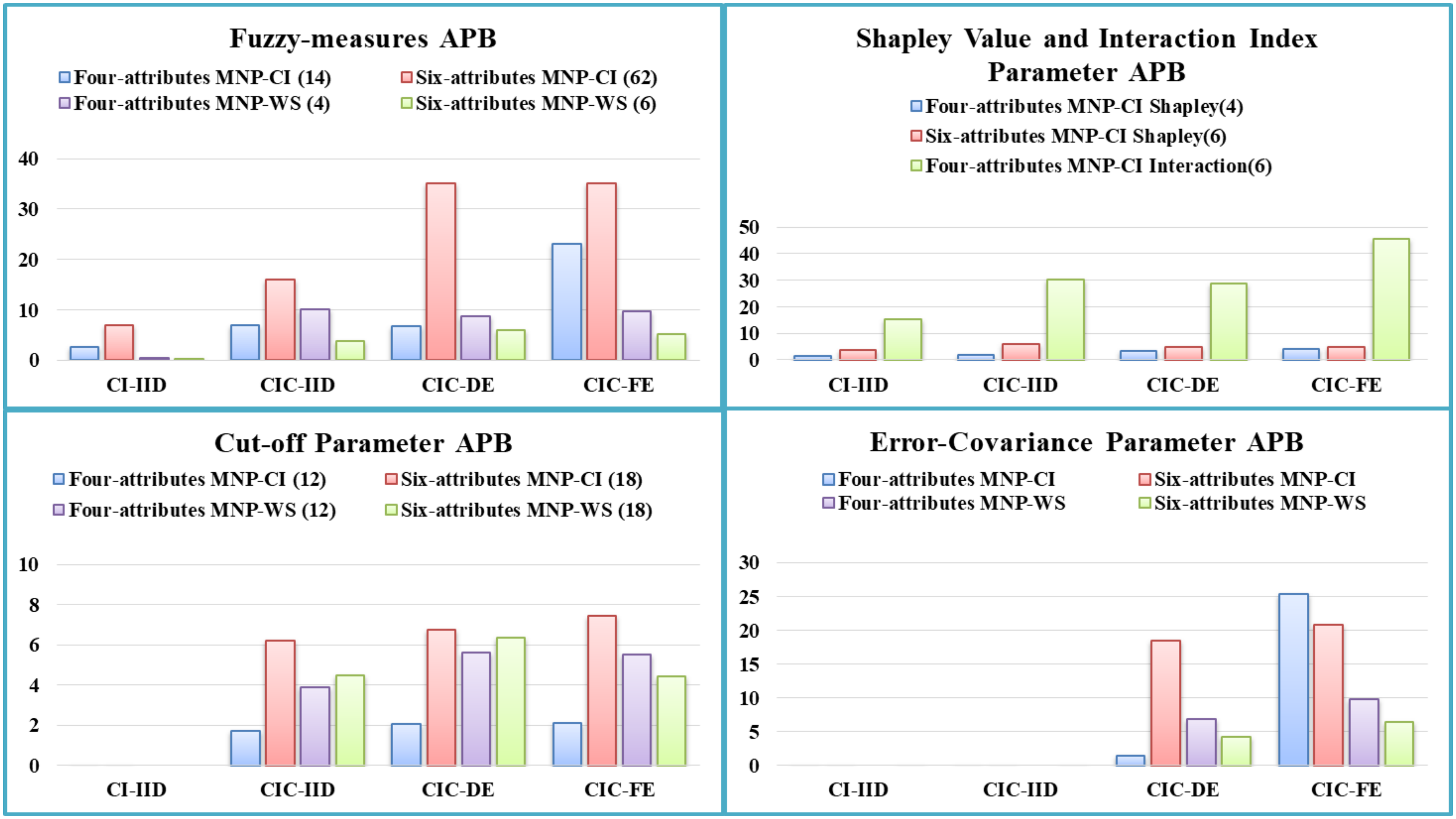

**Figure O.3:** Absolute percentage bias (APB) for various parameter groups (the number of parameters in parenthesis).
**(Note:** APB for interaction indices is not presented in case of MNP-CI with six attributes because true interaction indices are zero.)

We present additional results of the Monte Carlo study where we evaluate the generality of the MNP-CI model. In data generating process and estimation, full error covariance and attribute cut-offs are considered. MNP-CI and MNP-WS only differs in terms of aggregation function.

**Table O.4:** Change in probability when data generating process follows MNP-CI

| Variable | Quantile | MNP-CI Model | | | | | MNP-WS Model | | | | |
|---|---|---|---|---|---|---|---|---|---|---|---|
| | | Alt 1 | Alt 2 | Alt 3 | Alt 4 | Alt 5 | Alt 1 | Alt 2 | Alt 3 | Alt 4 | Alt 5 |
| 1 | 0.10 | -0.153 | -0.124 | -0.114 | -0.114 | -0.123 | -0.025 | -0.014 | -0.019 | -0.018 | -0.020 |
| 1 | 0.20 | -0.147 | -0.112 | -0.108 | -0.109 | -0.117 | -0.024 | -0.014 | -0.018 | -0.017 | -0.019 |
| 1 | 0.30 | -0.143 | -0.105 | -0.104 | -0.104 | -0.111 | -0.023 | -0.013 | -0.017 | -0.017 | -0.019 |
| 1 | 0.40 | -0.139 | -0.100 | -0.101 | -0.100 | -0.108 | -0.023 | -0.013 | -0.017 | -0.017 | -0.018 |
| 1 | 0.50 | -0.136 | -0.096 | -0.097 | -0.097 | -0.105 | -0.022 | -0.013 | -0.017 | -0.016 | -0.018 |
| 1 | 0.60 | -0.134 | -0.093 | -0.095 | -0.094 | -0.102 | -0.022 | -0.013 | -0.016 | -0.016 | -0.018 |
| 1 | 0.70 | -0.132 | -0.090 | -0.092 | -0.092 | -0.099 | -0.022 | -0.012 | -0.016 | -0.016 | -0.017 |
| 1 | 0.80 | -0.130 | -0.087 | -0.089 | -0.090 | -0.097 | -0.021 | -0.012 | -0.016 | -0.016 | -0.017 |
| 1 | 0.90 | -0.128 | -0.086 | -0.087 | -0.088 | -0.095 | -0.021 | -0.012 | -0.016 | -0.016 | -0.017 |
| 1 | 0.99 | -0.126 | -0.084 | -0.085 | -0.086 | -0.093 | -0.021 | -0.012 | -0.016 | -0.015 | -0.017 |
| 2 | 0.10 | -0.107 | -0.080 | -0.072 | -0.083 | -0.081 | -0.008 | -0.005 | -0.006 | -0.006 | -0.006 |
| 2 | 0.20 | -0.100 | -0.072 | -0.067 | -0.074 | -0.076 | -0.008 | -0.004 | -0.006 | -0.006 | -0.006 |
| 2 | 0.30 | -0.096 | -0.067 | -0.063 | -0.068 | -0.073 | -0.008 | -0.004 | -0.005 | -0.006 | -0.006 |
| 2 | 0.40 | -0.092 | -0.064 | -0.061 | -0.064 | -0.071 | -0.008 | -0.004 | -0.005 | -0.005 | -0.006 |
| 2 | 0.50 | -0.090 | -0.062 | -0.058 | -0.061 | -0.069 | -0.008 | -0.004 | -0.005 | -0.005 | -0.006 |
| 2 | 0.60 | -0.088 | -0.060 | -0.057 | -0.059 | -0.067 | -0.007 | -0.004 | -0.005 | -0.005 | -0.006 |
| 2 | 0.70 | -0.086 | -0.058 | -0.055 | -0.057 | -0.065 | -0.007 | -0.004 | -0.005 | -0.005 | -0.006 |
| 2 | 0.80 | -0.084 | -0.057 | -0.054 | -0.055 | -0.064 | -0.007 | -0.004 | -0.005 | -0.005 | -0.006 |
| 2 | 0.90 | -0.083 | -0.055 | -0.053 | -0.054 | -0.063 | -0.007 | -0.004 | -0.005 | -0.005 | -0.005 |
| 2 | 0.99 | -0.082 | -0.054 | -0.051 | -0.053 | -0.062 | -0.007 | -0.004 | -0.005 | -0.005 | -0.005 |
| 3 | 0.10 | -0.344 | -0.231 | -0.226 | -0.243 | -0.260 | -0.037 | -0.022 | -0.028 | -0.028 | -0.029 |
| 3 | 0.20 | -0.331 | -0.214 | -0.211 | -0.230 | -0.246 | -0.036 | -0.021 | -0.027 | -0.026 | -0.028 |
| 3 | 0.30 | -0.322 | -0.202 | -0.198 | -0.221 | -0.234 | -0.035 | -0.020 | -0.026 | -0.026 | -0.027 |
| 3 | 0.40 | -0.314 | -0.193 | -0.188 | -0.213 | -0.224 | -0.035 | -0.019 | -0.026 | -0.025 | -0.027 |
| 3 | 0.50 | -0.306 | -0.185 | -0.179 | -0.206 | -0.217 | -0.034 | -0.019 | -0.025 | -0.025 | -0.026 |
| 3 | 0.60 | -0.299 | -0.179 | -0.172 | -0.200 | -0.210 | -0.034 | -0.019 | -0.025 | -0.024 | -0.026 |
| 3 | 0.70 | -0.292 | -0.173 | -0.165 | -0.195 | -0.203 | -0.033 | -0.018 | -0.024 | -0.024 | -0.025 |
| 3 | 0.80 | -0.286 | -0.167 | -0.159 | -0.188 | -0.197 | -0.033 | -0.018 | -0.024 | -0.024 | -0.025 |
| 3 | 0.90 | -0.280 | -0.162 | -0.154 | -0.183 | -0.190 | -0.032 | -0.018 | -0.024 | -0.023 | -0.025 |
| 3 | 0.99 | -0.274 | -0.158 | -0.149 | -0.178 | -0.185 | -0.032 | -0.018 | -0.023 | -0.023 | -0.025 |
| 4 | 0.10 | -0.201 | -0.139 | -0.132 | -0.148 | -0.150 | -0.015 | -0.009 | -0.011 | -0.011 | -0.012 |
| 4 | 0.20 | -0.190 | -0.124 | -0.121 | -0.131 | -0.140 | -0.015 | -0.008 | -0.011 | -0.011 | -0.012 |
| 4 | 0.30 | -0.180 | -0.117 | -0.115 | -0.124 | -0.132 | -0.014 | -0.008 | -0.010 | -0.010 | -0.011 |
| 4 | 0.40 | -0.173 | -0.111 | -0.110 | -0.119 | -0.127 | -0.014 | -0.008 | -0.010 | -0.010 | -0.011 |
| 4 | 0.50 | -0.167 | -0.107 | -0.106 | -0.115 | -0.123 | -0.014 | -0.008 | -0.010 | -0.010 | -0.011 |
| 4 | 0.60 | -0.162 | -0.104 | -0.103 | -0.112 | -0.120 | -0.013 | -0.007 | -0.010 | -0.010 | -0.011 |
| 4 | 0.70 | -0.158 | -0.101 | -0.100 | -0.109 | -0.117 | -0.013 | -0.007 | -0.009 | -0.010 | -0.010 |
| 4 | 0.80 | -0.154 | -0.098 | -0.098 | -0.106 | -0.114 | -0.013 | -0.007 | -0.009 | -0.010 | -0.010 |
| 4 | 0.90 | -0.151 | -0.096 | -0.096 | -0.104 | -0.112 | -0.013 | -0.007 | -0.009 | -0.009 | -0.010 |
| 4 | 0.99 | -0.148 | -0.094 | -0.094 | -0.102 | -0.110 | -0.013 | -0.007 | -0.009 | -0.009 | -0.010 |

**Table O.5:** Change in probability when data generating process follows MNP-WS

| Variable | Quantile | MNP-CI Model | | | | | MNP-WS Model | | | | |
|---|---|---|---|---|---|---|---|---|---|---|---|
| | | Alt 1 | Alt 2 | Alt 3 | Alt 4 | Alt 5 | Alt 1 | Alt 2 | Alt 3 | Alt 4 | Alt 5 |
| 1 | 0.10 | -0.076 | -0.050 | -0.050 | -0.048 | -0.053 | -0.053 | -0.034 | -0.038 | -0.042 | -0.043 |
| 1 | 0.20 | -0.065 | -0.044 | -0.043 | -0.043 | -0.046 | -0.051 | -0.032 | -0.036 | -0.041 | -0.041 |
| 1 | 0.30 | -0.058 | -0.041 | -0.038 | -0.039 | -0.042 | -0.050 | -0.031 | -0.035 | -0.039 | -0.039 |
| 1 | 0.40 | -0.054 | -0.039 | -0.036 | -0.037 | -0.039 | -0.049 | -0.030 | -0.034 | -0.038 | -0.039 |
| 1 | 0.50 | -0.051 | -0.038 | -0.034 | -0.036 | -0.038 | -0.049 | -0.029 | -0.033 | -0.038 | -0.038 |
| 1 | 0.60 | -0.048 | -0.037 | -0.033 | -0.035 | -0.037 | -0.048 | -0.028 | -0.033 | -0.037 | -0.037 |
| 1 | 0.70 | -0.047 | -0.036 | -0.032 | -0.034 | -0.036 | -0.048 | -0.028 | -0.032 | -0.036 | -0.037 |
| 1 | 0.80 | -0.046 | -0.035 | -0.030 | -0.033 | -0.035 | -0.047 | -0.027 | -0.032 | -0.036 | -0.036 |
| 1 | 0.90 | -0.045 | -0.035 | -0.030 | -0.033 | -0.034 | -0.046 | -0.027 | -0.031 | -0.035 | -0.036 |
| 1 | 0.99 | -0.044 | -0.034 | -0.029 | -0.032 | -0.034 | -0.046 | -0.026 | -0.031 | -0.035 | -0.035 |
| 2 | 0.10 | -0.052 | -0.039 | -0.035 | -0.037 | -0.040 | -0.048 | -0.030 | -0.035 | -0.036 | -0.038 |
| 2 | 0.20 | -0.049 | -0.035 | -0.031 | -0.034 | -0.038 | -0.047 | -0.029 | -0.033 | -0.035 | -0.037 |
| 2 | 0.30 | -0.046 | -0.033 | -0.030 | -0.032 | -0.036 | -0.047 | -0.027 | -0.032 | -0.034 | -0.036 |
| 2 | 0.40 | -0.043 | -0.031 | -0.028 | -0.030 | -0.034 | -0.046 | -0.027 | -0.031 | -0.033 | -0.035 |
| 2 | 0.50 | -0.042 | -0.030 | -0.027 | -0.029 | -0.032 | -0.045 | -0.026 | -0.030 | -0.032 | -0.035 |
| 2 | 0.60 | -0.041 | -0.029 | -0.026 | -0.028 | -0.031 | -0.045 | -0.026 | -0.030 | -0.032 | -0.034 |
| 2 | 0.70 | -0.039 | -0.027 | -0.025 | -0.027 | -0.030 | -0.044 | -0.025 | -0.029 | -0.031 | -0.034 |
| 2 | 0.80 | -0.039 | -0.026 | -0.024 | -0.026 | -0.030 | -0.044 | -0.024 | -0.029 | -0.031 | -0.033 |
| 2 | 0.90 | -0.038 | -0.026 | -0.024 | -0.025 | -0.029 | -0.044 | -0.024 | -0.029 | -0.031 | -0.033 |
| 2 | 0.99 | -0.037 | -0.025 | -0.023 | -0.025 | -0.028 | -0.043 | -0.024 | -0.028 | -0.030 | -0.033 |
| 3 | 0.10 | -0.119 | -0.078 | -0.074 | -0.079 | -0.083 | -0.122 | -0.067 | -0.082 | -0.090 | -0.092 |
| 3 | 0.20 | -0.110 | -0.073 | -0.067 | -0.072 | -0.077 | -0.118 | -0.063 | -0.078 | -0.085 | -0.088 |
| 3 | 0.30 | -0.103 | -0.070 | -0.063 | -0.068 | -0.072 | -0.115 | -0.060 | -0.075 | -0.082 | -0.085 |
| 3 | 0.40 | -0.100 | -0.067 | -0.061 | -0.066 | -0.070 | -0.113 | -0.058 | -0.072 | -0.080 | -0.083 |
| 3 | 0.50 | -0.096 | -0.065 | -0.059 | -0.064 | -0.068 | -0.111 | -0.056 | -0.071 | -0.078 | -0.081 |
| 3 | 0.60 | -0.094 | -0.063 | -0.057 | -0.062 | -0.066 | -0.109 | -0.055 | -0.069 | -0.076 | -0.080 |
| 3 | 0.70 | -0.092 | -0.062 | -0.056 | -0.061 | -0.065 | -0.108 | -0.054 | -0.068 | -0.075 | -0.079 |
| 3 | 0.80 | -0.090 | -0.06 | -0.055 | -0.059 | -0.063 | -0.106 | -0.053 | -0.066 | -0.073 | -0.078 |
| 3 | 0.90 | -0.088 | -0.059 | -0.053 | -0.058 | -0.062 | -0.105 | -0.052 | -0.065 | -0.072 | -0.077 |
| 3 | 0.99 | -0.087 | -0.058 | -0.053 | -0.057 | -0.062 | -0.104 | -0.051 | -0.064 | -0.071 | -0.076 |
| 4 | 0.10 | -0.066 | -0.047 | -0.040 | -0.047 | -0.051 | -0.046 | -0.028 | -0.032 | -0.035 | -0.036 |
| 4 | 0.20 | -0.057 | -0.041 | -0.036 | -0.041 | -0.044 | -0.045 | -0.026 | -0.030 | -0.033 | -0.034 |
| 4 | 0.30 | -0.052 | -0.037 | -0.033 | -0.037 | -0.040 | -0.043 | -0.025 | -0.029 | -0.032 | -0.033 |
| 4 | 0.40 | -0.048 | -0.035 | -0.031 | -0.035 | -0.037 | -0.042 | -0.024 | -0.028 | -0.031 | -0.032 |
| 4 | 0.50 | -0.046 | -0.033 | -0.029 | -0.033 | -0.035 | -0.042 | -0.023 | -0.027 | -0.030 | -0.032 |
| 4 | 0.60 | -0.045 | -0.031 | -0.028 | -0.031 | -0.033 | -0.041 | -0.023 | -0.027 | -0.030 | -0.031 |
| 4 | 0.70 | -0.043 | -0.030 | -0.027 | -0.030 | -0.032 | -0.040 | -0.022 | -0.026 | -0.029 | -0.031 |
| 4 | 0.80 | -0.042 | -0.029 | -0.026 | -0.029 | -0.031 | -0.040 | -0.022 | -0.026 | -0.029 | -0.030 |
| 4 | 0.90 | -0.041 | -0.028 | -0.025 | -0.028 | -0.030 | -0.039 | -0.021 | -0.025 | -0.028 | -0.030 |
| 4 | 0.99 | -0.040 | -0.027 | -0.025 | -0.028 | -0.029 | -0.039 | -0.021 | -0.025 | -0.028 | -0.029 |

## S.7 Additional Results of the Empirical Study

**Table O.6:** Choquet integral fuzzy-measure estimates (T-statistics in parenthesis)

| Variables | CI-NAC | CI-CAC | CI-GAC | CI-AGAC | ACI-AGAC | |
|---|---|---|---|---|---|---|
| | **All modes** | **All modes** | **All modes** | **All modes** | **Current mode** | **Uber and Uberpool** |
| IVTT/Km | 0.068 (1.6) | 0.000 (0.0) | 0.000 (0.0) | 0.000 (0.0) | 0.000 (0.0) | 0.060 (0.6) |
| OVTT/Km | 0.173 (3.5) | 0.143 (3.9) | 0.076 (1.9) | 0.000 (0.0) | 0.003 (0.1) | 0.000 (0.0) |
| Cost/Km | 0.427 (8.9) | 0.318 (6.8) | 0.361 (7.7) | 0.297 (7.0) | 0.302 (6.9) | 0.270 (2.3) |
| IVTT/Km, OVTT/Km | 0.381 (7.5) | 0.353 (4.5) | 0.366 (3.4) | 0.468 (7.5) | 0.515 (7.2) | 0.427 (4.1) |
| IVTT/Km, Cost/Km | 0.577 (10.4) | 0.668 (6.8) | 0.725 (6.4) | 0.578 (9.2) | 0.569 (8.5) | 0.683 (8.4) |
| OVTT/Km, Cost/Km | 0.626 (10.3) | 0.928 (10.0) | 0.984 (13.3) | 1.000 (13.7) | 1.000 (13.4) | 0.862 (8.0) |
| IVTT/Km, OVTT/Km, Cost/Km | 1.000 (16.5) | 1.000 (9.1) | 1.000 (9.7) | 1.000 (12.9) | 1.000 (12.6) | 1.000 (11.5) |

**Note:** IVTT and OVTT imply in-vehicle and out-of-vehicle travel time.

**Table O.7:** Attribute cut-off heterogeneity in CI-AGAC model for current travel mode (T-statistic in parenthesis)

| **Explanatory variables** | | **IVTT/Km** | | **OVTT/Km** | | **Cost/Km** | |
|---|---|---|---|---|---|---|---|
| | | **Lower Cut-off** | **Upper Cut-off** | **Lower Cut-off** | **Upper Cut-off** | **Lower Cut-off** | **Upper Cut-off** |
| | Constant | 1.11 (18.5) | -0.7 (-0.9) | -0.6 (-3.6) | 2.11 (31.9) | -1.27 (-7.3) | 0.76 (13.3) |
| Household income * travel distance (in USD * km) [Base: >125K] | (<=50K) * distance | | | | | | -0.58 (-6.5) |
| | (> 50K & <=125K) * distance | | | | | | -0.36 (-5.1) |
| Distance to Bus stop (in km) [Base: ≤ 0.5] | (> 0.5 & ≤ 1) | | | | -0.53 (-3.9) | | |
| | (> 1 & ≤ 2) | | | | -2.75 (-5.6) | | |
| | (> 2) | | | | -2.75 (-5.6) | | |
| Distance to subway (in km) [Base: ≤ 0.5] | (> 0.5 & ≤ 1) | | | | -0.20 (-1.8) | | |
| | (> 1 & ≤ 2) | | | | -0.63 (-3.7) | | |
| | (> 2) | | | | -0.95 (-6.3) | | |
| Male | | | | | | | |
| Years since owing a driver's license | | | | | | | |
| Age (in years) [Base: 23 – 38] | Age (7 - 22) | | | | 0.23 (1.8) | | |
| | Age (39 - 54) | | | | | | |
| | Age (55 - 73) | | | | -0.83 (-5.9) | | |

**Table O.8:** Attribute cut-off heterogeneity in CI-AGAC model for Uber (T-statistic in parenthesis)

| **Explanatory variables** | | **IVTT/Km** | | **OVTT/Km** | | **Cost/Km** | |
|---|---|---|---|---|---|---|---|
| | | **Lower Cut-off** | **Upper Cut-off** | **Lower Cut-off** | **Upper Cut-off** | **Lower Cut-off** | **Upper Cut-off** |
| | Constant | -0.13<br>(-0.2) | 1.46<br>(5.2) | 0.66<br>(25.7) | 0.95<br>(5.7) | -3.13<br>(-1.2) | 1.01<br>(11.2) |
| Household income * travel distance (in USD * km) [Base: >125K] | (<=50K) * distance | | | | | | |
| | (> 50K & <=125K) * distance | | | | | | |
| Distance to Bus stop (in km) [Base: ≤ 0.5] | (> 0.5 & ≤ 1) | | | | -0.53<br>(-3.9) | | |
| | (> 1 & ≤ 2) | | | | -2.75<br>(-5.6) | | |
| | (> 2) | | | | -2.75<br>(-5.6) | | |
| Distance to subway (in km) [Base: ≤ 0.5] | (> 0.5 & ≤ 1) | | | | -0.20<br>(-1.8) | | |
| | (> 1 & ≤ 2) | | | | -0.63<br>(-3.7) | | |
| | (> 2) | | | | -0.95<br>(-6.3) | | |
| Male | | | | | | | |
| Years since owing a driver's license | | | | | | | |
| Age (in years) [Base: 23 – 38] | Age (7 - 22) | | | | 0.23<br>(1.8) | | |
| | Age (39 - 54) | | | | | | -0.5<br>(-4.6) |
| | Age (55 - 73) | | | | -0.83<br>(-5.9) | | -3.24<br>(-1.4) |

**Table O.9:** Attribute cut-off heterogeneity in CI-AGAC model for Uberpool (T-statistic in parenthesis)

| **Explanatory variables** | | **IVTT/Km** | | **OVTT/Km** | | **Cost/Km** | |
|---|---|---|---|---|---|---|---|
| | | **Lower Cut-off** | **Upper Cut-off** | **Lower Cut-off** | **Upper Cut-off** | **Lower Cut-off** | **Upper Cut-off** |
| | Constant | -0.64 (-0.3) | 2.07 (13.8) | -1.11 (-6.4) | 1.37 (12.7) | -2.31 (-5.8) | 1.23 (13.2) |
| Household income * travel distance (in USD * km) [Base: >125K] | (<=50K) * distance | | | | | | -0.61 (-6.9) |
| | (> 50K & <=125K) * distance | | | | | | -0.47 (-5.6) |
| Distance to Bus stop (in km) [Base: ≤ 0.5] | (> 0.5 & ≤ 1) | | | | -0.53 (-3.9) | | |
| | (> 1 & ≤ 2) | | | | -2.75 (-5.6) | | |
| | (> 2) | | | | -2.75 (-5.6) | | |
| Distance to subway (in km) [Base: ≤ 0.5] | (> 0.5 & ≤ 1) | | | | -0.20 (-1.8) | | |
| | (> 1 & ≤ 2) | | | | -0.63 (-3.7) | | |
| | (> 2) | | | | -0.95 (-6.3) | | |
| Male | | | | | | | |
| Years since owing a driver's license | | | | | | | |
| Age (in years) [Base: 23 – 38] | Age (7 - 22) | | | | 0.23 (1.8) | | 0.34 (2.7) |
| | Age (39 - 54) | | | | | | -0.59 (-5.2) |
| | Age (55 - 73) | | | | -0.83 (-5.9) | | -1.24 (-6.7) |

**Table O.10:** Attribute cut-off heterogeneity in ACI-AGAC model for current travel mode (T-statistic in parenthesis)

| **Explanatory variables** | | **IVTT/Km** | | **OVTT/Km** | | **Cost/Km** | |
|---|---|---|---|---|---|---|---|
| | | **Lower Cut-off** | **Upper Cut-off** | **Lower Cut-off** | **Upper Cut-off** | **Lower Cut-off** | **Upper Cut-off** |
| | Constant | 1.11 (20.5) | -0.71 (-0.9) | -0.59 (-3.0) | 2.1 (31.1) | -1.29 (-6.7) | 0.74 (12.2) |
| Household income * travel distance (in USD * km) [Base: >125K] | (<=50K) * distance | | | | | | -0.55 (-6.7) |
| | (> 50K & <=125K) * distance | | | | | | -0.35 (-4.8) |
| Distance to Bus stop (in km) [Base: ≤ 0.5] | (> 0.5 & ≤ 1) | | | | -0.53 (-3.9) | | |
| | (> 1 & ≤ 2) | | | | -2.79 (-4.2) | | |
| | (> 2) | | | | -2.79 (-4.2) | | |
| Distance to subway (in km) [Base: ≤ 0.5] | (> 0.5 & ≤ 1) | | | | -0.19 (-1.6) | | |
| | (> 1 & ≤ 2) | | | | -0.61 (-3.4) | | |
| | (> 2) | | | | -0.91 (-6.0) | | |
| Male | | | | | | | |
| Years since owing a driver's license | | | | | | | |
| Age (in years) [Base: 23 – 38] | Age (7 - 22) | | | | 0.23 (1.7) | | |
| | Age (39 - 54) | | | | | | |
| | Age (55 - 73) | | | | -0.82 (-5.5) | | |

**Table O.11:** Attribute cut-off heterogeneity in ACI-AGAC model for Uber (T-statistic in parenthesis)

| Explanatory variables | | IVTT/Km | | OVTT/Km | | Cost/Km | |
|---|---|---|---|---|---|---|---|
| | | Lower Cut-off | Upper Cut-off | Lower Cut-off | Upper Cut-off | Lower Cut-off | Upper Cut-off |
| | Constant | -0.16 (-0.2) | 1.47 (5.3) | 0.66 (8.6) | 0.95 (4.4) | -3.15 (-1.2) | 1.06 (10.0) |
| Household income * travel distance (in USD * km) [Base: >125K] | (<=50K) * distance | | | | | | |
| | (> 50K & <=125K) * distance | | | | | | |
| Distance to Bus stop (in km) [Base: ≤ 0.5] | (> 0.5 & ≤ 1) | | | | -0.53 (-3.9) | | |
| | (> 1 & ≤ 2) | | | | -2.79 (-4.2) | | |
| | (> 2) | | | | -2.79 (-4.2) | | |
| Distance to subway (in km) [Base: ≤ 0.5] | (> 0.5 & ≤ 1) | | | | -0.19 (-1.6) | | |
| | (> 1 & ≤ 2) | | | | -0.61 (-3.4) | | |
| | (> 2) | | | | -0.91 (-6.0) | | |
| Male | | | | | | | |
| Years since owing a driver's license | | | | | | | |
| Age (in years) [Base: 23 – 38] | Age (7 - 22) | | | | 0.23 (1.7) | | |
| | Age (39 - 54) | | | | | | -0.55 (-4.2) |
| | Age (55 - 73) | | | | -0.82 (-5.5) | | -3.26 (-2.4) |

**Table O.12:** Attribute cut-off heterogeneity in ACI-AGAC model for Uberpool (T-statistic in parenthesis)

| **Explanatory variables** | | **IVTT/Km** | | **OVTT/Km** | | **Cost/Km** | |
|---|---|---|---|---|---|---|---|
| | | **Lower Cut-off** | **Upper Cut-off** | **Lower Cut-off** | **Upper Cut-off** | **Lower Cut-off** | **Upper Cut-off** |
| | Constant | -0.65 (-0.2) | 2.06 (12.6) | -1.12 (-5.1) | 1.34 (10.3) | -2.28 (-4.0) | 1.26 (11.3) |
| Household income * travel distance (in USD * km) [Base: >125K] | (<=50K) * distance | | | | | | -0.49 (-4.8) |
| | (> 50K & <=125K) * distance | | | | | | -0.47 (-4.8) |
| Distance to Bus stop (in km) [Base: ≤ 0.5] | (> 0.5 & ≤ 1) | | | | -0.53 (-3.9) | | |
| | (> 1 & ≤ 2) | | | | -2.79 (-4.2) | | |
| | (> 2) | | | | -2.79 (-4.2) | | |
| Distance to subway (in km) [Base: ≤ 0.5] | (> 0.5 & ≤ 1) | | | | -0.19 (-1.6) | | |
| | (> 1 & ≤ 2) | | | | -0.61 (-3.4) | | |
| | (> 2) | | | | -0.91 (-6.0) | | |
| Male | | | | | | | -0.37 (-3.8) |
| Years since owing a driver's license | | | | | | | |
| Age (in years) [Base: 23 – 38] | Age (7 - 22) | | | | 0.23 (1.7) | | 0.36 (2.4) |
| | Age (39 - 54) | | | | | | -0.61 (-5) |
| | Age (55 - 73) | | | | -0.82 (-5.5) | | -1.27 (-6.2) |

**Table O.13:** Distribution of attribute cut-off for current travel mode in CI-AGAC model

| Percentile | IVTT/Km | | OVTT/Km | | Cost/Km | |
|---|---|---|---|---|---|---|
| | Lower Cut-off | Upper Cut-off | Lower Cut-off | Upper Cut-off | Lower Cut-off | Upper Cut-off |
| 10 | 3.02 | 3.52 | 0.55 | 1.36 | 0.28 | 1.18 |
| 20 | 3.02 | 3.52 | 0.55 | 3.74 | 0.28 | 1.48 |
| 30 | 3.02 | 3.52 | 0.55 | 4.12 | 0.28 | 1.73 |
| 40 | 3.02 | 3.52 | 0.55 | 5.49 | 0.28 | 1.89 |
| 50 | 3.02 | 3.52 | 0.55 | 7.28 | 0.28 | 2.01 |
| 60 | 3.02 | 3.52 | 0.55 | 8.76 | 0.28 | 2.08 |
| 70 | 3.02 | 3.52 | 0.55 | 8.76 | 0.28 | 2.21 |
| 80 | 3.02 | 3.52 | 0.55 | 8.76 | 0.28 | 2.43 |
| 90 | 3.02 | 3.52 | 0.55 | 8.76 | 0.28 | 2.43 |
| 100 | 3.02 | 3.52 | 0.55 | 10.83 | 0.28 | 2.43 |

**Table O.14:** Distribution of attribute cut-off for Uber in CI-AGAC model

| Percentile | IVTT/Km | | OVTT/Km | | Cost/Km | |
|---|---|---|---|---|---|---|
| | Lower Cut-off | Upper Cut-off | Lower Cut-off | Upper Cut-off | Lower Cut-off | Upper Cut-off |
| 10 | 0.88 | 5.19 | 1.93 | 2.19 | 0.04 | 0.2 |
| 20 | 0.88 | 5.19 | 1.93 | 2.94 | 0.04 | 1.71 |
| 30 | 0.88 | 5.19 | 1.93 | 3.06 | 0.04 | 2.41 |
| 40 | 0.88 | 5.19 | 1.93 | 3.49 | 0.04 | 2.79 |
| 50 | 0.88 | 5.19 | 1.93 | 4.06 | 0.04 | 2.79 |
| 60 | 0.88 | 5.19 | 1.93 | 4.52 | 0.04 | 2.79 |
| 70 | 0.88 | 5.19 | 1.93 | 4.52 | 0.04 | 2.79 |
| 80 | 0.88 | 5.19 | 1.93 | 4.52 | 0.04 | 2.79 |
| 90 | 0.88 | 5.19 | 1.93 | 4.52 | 0.04 | 3.95 |
| 100 | 0.88 | 5.19 | 1.93 | 5.18 | 0.04 | 3.95 |

**Table O.15:** Distribution of attribute cut-off for Uberpool in CI-AGAC model

| Percentile | IVTT/Km | | OVTT/Km | | Cost/Km | |
|---|---|---|---|---|---|---|
| | Lower Cut-off | Upper Cut-off | Lower Cut-off | Upper Cut-off | Lower Cut-off | Upper Cut-off |
| 10 | 0.53 | 8.45 | 0.33 | 0.72 | 0.1 | 0.76 |
| 20 | 0.53 | 8.45 | 0.33 | 1.86 | 0.1 | 1.09 |
| 30 | 0.53 | 8.45 | 0.33 | 2.04 | 0.1 | 1.37 |
| 40 | 0.53 | 8.45 | 0.33 | 2.7 | 0.1 | 1.61 |
| 50 | 0.53 | 8.45 | 0.33 | 3.56 | 0.1 | 1.97 |
| 60 | 0.53 | 8.45 | 0.33 | 4.27 | 0.1 | 2.24 |
| 70 | 0.53 | 8.45 | 0.33 | 4.27 | 0.1 | 2.56 |
| 80 | 0.53 | 8.45 | 0.33 | 4.27 | 0.1 | 2.95 |
| 90 | 0.53 | 8.45 | 0.33 | 4.27 | 0.1 | 3.52 |
| 100 | 0.53 | 8.45 | 0.33 | 5.27 | 0.1 | 3.52 |

**Table O.16:** Distribution of attribute cut-off for current travel mode in ACI-AGAC model

| Percentile | IVTT/Km | | OVTT/Km | | Cost/Km | |
|---|---|---|---|---|---|---|
| | Lower Cut-off | Upper Cut-off | Lower Cut-off | Upper Cut-off | Lower Cut-off | Upper Cut-off |
| 10 | 3.03 | 3.52 | 0.55 | 1.40 | 0.28 | 1.20 |
| 20 | 3.03 | 3.52 | 0.55 | 3.83 | 0.28 | 1.49 |
| 30 | 3.03 | 3.52 | 0.55 | 4.12 | 0.28 | 1.71 |
| 40 | 3.03 | 3.52 | 0.55 | 5.50 | 0.28 | 1.86 |
| 50 | 3.03 | 3.52 | 0.55 | 7.29 | 0.28 | 1.98 |
| 60 | 3.03 | 3.52 | 0.55 | 8.69 | 0.28 | 2.06 |
| 70 | 3.03 | 3.52 | 0.55 | 8.69 | 0.28 | 2.16 |
| 80 | 3.03 | 3.52 | 0.55 | 8.69 | 0.28 | 2.37 |
| 90 | 3.03 | 3.52 | 0.55 | 8.69 | 0.28 | 2.37 |
| 100 | 3.03 | 3.52 | 0.55 | 10.74 | 0.28 | 2.37 |

**Table O.17:** Distribution of attribute cut-off for Uber in ACI-AGAC model

| Percentile | IVTT/Km | | OVTT/Km | | Cost/Km | |
|---|---|---|---|---|---|---|
| | Lower Cut-off | Upper Cut-off | Lower Cut-off | Upper Cut-off | Lower Cut-off | Upper Cut-off |
| 10 | 0.85 | 5.19 | 1.93 | 2.20 | 0.04 | 0.20 |
| 20 | 0.85 | 5.19 | 1.93 | 2.98 | 0.04 | 1.71 |
| 30 | 0.85 | 5.19 | 1.93 | 3.07 | 0.04 | 2.47 |
| 40 | 0.85 | 5.19 | 1.93 | 3.51 | 0.04 | 2.93 |
| 50 | 0.85 | 5.19 | 1.93 | 4.08 | 0.04 | 2.93 |
| 60 | 0.85 | 5.19 | 1.93 | 4.52 | 0.04 | 2.93 |
| 70 | 0.85 | 5.19 | 1.93 | 4.52 | 0.04 | 2.93 |
| 80 | 0.85 | 5.19 | 1.93 | 4.52 | 0.04 | 2.93 |
| 90 | 0.85 | 5.19 | 1.93 | 4.52 | 0.04 | 4.24 |
| 100 | 0.85 | 5.19 | 1.93 | 5.18 | 0.04 | 4.24 |

**Table O.18:** Distribution of attribute cut-off for Uberpool in ACI-AGAC model

| Percentile | IVTT/Km | | OVTT/Km | | Cost/Km | |
|---|---|---|---|---|---|---|
| | Lower Cut-off | Upper Cut-off | Lower Cut-off | Upper Cut-off | Lower Cut-off | Upper Cut-off |
| 10 | 0.52 | 8.37 | 0.33 | 0.73 | 0.10 | 0.79 |
| 20 | 0.52 | 8.37 | 0.33 | 1.87 | 0.10 | 1.10 |
| 30 | 0.52 | 8.37 | 0.33 | 2.01 | 0.10 | 1.43 |
| 40 | 0.52 | 8.37 | 0.33 | 2.66 | 0.10 | 1.78 |
| 50 | 0.52 | 8.37 | 0.33 | 3.50 | 0.10 | 2.03 |
| 60 | 0.52 | 8.37 | 0.33 | 4.15 | 0.10 | 2.42 |
| 70 | 0.52 | 8.37 | 0.33 | 4.15 | 0.10 | 2.77 |
| 80 | 0.52 | 8.37 | 0.33 | 4.15 | 0.10 | 3.16 |
| 90 | 0.52 | 8.37 | 0.33 | 4.15 | 0.10 | 3.64 |
| 100 | 0.52 | 8.37 | 0.33 | 5.12 | 0.10 | 3.64 |

**Table O.19:** MNP-WS estimates with and without interaction (T-statistic in parenthesis)

| Variables | MNP-WS with no-interactions and no attribute cut-offs | | | MNP-WS with interactions and no attribute cut-offs | | |
|---|---|---|---|---|---|---|
| | **Current mode** | **Uber** | **Uberpool** | **Current mode** | **Uber** | **Uberpool** |
| Constant | | -0.28 (-6.0) | -0.59 (-8.2) | | -0.37 (-7.5) | -0.68 (-9.7) |
| Electric | | -0.08 (-2.3) | -0.09 (-2.5) | | -0.06 (-1.6) | -0.07 (-2.0) |
| Automated | | -0.14 (-4.1) | -0.10 (-2.9) | | -0.13 (-3.6) | -0.10 (-2.7) |
| IVTT/Km | -0.05 (-7.9) | | | -0.08 (-10.2) | | |
| OVTT/Km | -0.05 (-13.4) | | | -0.11 (-15.2) | | |
| Cost/Km | -0.16 (-26.9) | | | -0.33 (-21.8) | | |
| IVTT/Km, OVTT/Km | | | | 0.003 (7.5) | | |
| IVTT/Km, Cost/Km | | | | 0.008 (14.2) | | |
| OVTT/Km, Cost/Km | | | | 0.012 (16.0) | | |
| IVTT/Km, OVTT/Km, Cost/Km | | | | -0.0001 (-10.6) | | |
| Error-covariance | $\begin{bmatrix} 1.00 \text{ (fixed)} & \\ 0.422 \text{ (3.8)} & 1.108 \text{ (7.8)} \end{bmatrix}$ | | | $\begin{bmatrix} 1.00 \text{ (fixed)} & \\ 0.179 \text{ (1.6)} & 1.020 \text{ (6.9)} \end{bmatrix}$ | | |